\documentclass[9pt]{elife} 
\usepackage{bm}
\usepackage{multicol}
\usepackage{multirow}
\usepackage{wrapfig}

\usepackage[utf8]{inputenc} 
\usepackage[T1]{fontenc}    
\usepackage{hyperref}       
\usepackage{url}            
\usepackage{booktabs}       
\usepackage{amsmath}
\usepackage{nicefrac}       
\usepackage{microtype}      
\usepackage{hhline}
\usepackage{graphicx}
\usepackage[font=small,labelfont=bf,tableposition=top]{caption}
\DeclareCaptionLabelFormat{andtable}{#1~#2  \&  \tablename~\thetable}
\usepackage[english]{babel}
\usepackage[english]{babel}
\usepackage{float}
\usepackage{wrapfig}
\usepackage{amsthm}
\usepackage{listings}
\usepackage{courier}
\definecolor{codegreen}{rgb}{0,0.6,0}
\definecolor{codegray}{rgb}{0.5,0.5,0.5}
\definecolor{codepurple}{rgb}{0.58,0,0.82}
\definecolor{backcolour}{rgb}{0.95,0.95,0.92}
\usepackage{soul}
\usepackage{subcaption}
\sethlcolor{white}
\usepackage{xurl}   
\usepackage{siunitx}  
\DeclareSIUnit\angstrom{\text{Å}}
\DeclareSIUnit\atmosphere{atm}
\DeclareSIUnit\calorie{cal}

\lstdefinestyle{mystyle}{
    float=tp,
    floatplacement=tbp,
    commentstyle=\color{codegreen},
    keywordstyle=\color{magenta},
    numberstyle=\tiny\color{codegray},
    stringstyle=\color{codepurple},
    basicstyle=\ttfamily\small,
    breakatwhitespace=false,         
    breaklines=true,                 
    captionpos=b,                    
    keepspaces=true,                 
    numbers=left,                    
    numbersep=5pt,                  
    showspaces=false,                
    showstringspaces=false,
    showtabs=false,                  
    tabsize=2,
    frame=tb,
}

\newfloat{lstfloat}{t}{lop}
\floatname{lstfloat}{Listing}

\title{Machine-learned molecular mechanics force fields from large-scale quantum chemical data}

\author[1, 2]{Kenichiro~Takaba~(ORCID:~\href{https://orcid.org/0000-0002-2481-8830}{0000-0002-2481-8830})}
\author[1]{Iván~Pulido~(ORCID:~\href{http://orcid.org/0000-0002-7178-8136}{0000-0002-7178-8136})}
\author[3]{Pavan~Kumar~Behara~(ORCID:~\href{https://orcid.org/0000-0001-6583-2148}{0000-0001-6583-2148})}
\author[4]{Chapin~E.~Cavender~(ORCID:~\href{https://orcid.org/0000-0002-5899-7953}{0000-0002-5899-7953})}
\author[5]{Anika~J.~Friedman~(ORCID:~\href{https://orcid.org/0000-0002-5427-2779}{0000-0002-5427-2779})}
\author[1]{Michael~M.~Henry~(ORCID:~\href{https://orcid.org/0000-0002-3870-9993}{0000-0002-3870-9993})}
\author[6]{Hugo~MacDermott-Opeskin~(ORCID:\href{https://orcid.org/0000-0002-7393-7457}{0000-0002-7393-7457})}
\author[1]{Christopher~R.~Iacovella~(ORCID:\href{https://orcid.org/0000-0003-0557-0427}{0000-0003-0557-0427})}
\author[7, 1]{Arnav~M.~Nagle~(ORCID:~\href{https://orcid.org/0009-0002-6749-4917}{0009-0002-6749-4917})}
\author[1, 9]{Alexander~Matthew~Payne~(ORCID:~\href{https://orcid.org/0000-0003-0947-0191}{0000-0003-0947-0191})}
\author[5]{Michael~R.~Shirts~(ORCID:~\href{https://orcid.org/0000-0003-3249-1097}{0000-0003-3249-1097})}
\author[8]{David~L.~Mobley~(ORCID:~\href{https://orcid.org/0000-0002-1083-5533}{0000-0002-1083-5533})}
\author[1]{John~D.~Chodera~(ORCID:~\href{http://orcid.org/0000-0003-0542-119X}{0000-0003-0542-119X})}
\author[10, 1]{Yuanqing~Wang~(ORCID:~\href{https://orcid.org/0000-0003-4403-2015}{0000-0003-4403-2015})}

\affil[1]{Computational and Systems Biology Program, Sloan Kettering Institute, Memorial Sloan Kettering Cancer Center, New York, N.Y. 10065, United States}
\affil[2]{Pharmaceuticals Research Center, Advanced Drug Discovery, Asahi Kasei Pharma Corporation, Shizuoka 410-2321, Japan}
\affil[3]{Center for Neurotherapeutics, Department of Pathology and Laboratory Medicine, University of California, Irvine, CA 92697, United States}
\affil[4]{Skaggs School of Pharmacy and Pharmaceutical Sciences, University of California San Diego, 9500 Gilman Drive, La Jolla, CA, 92093, United States}
\affil[5]{Department of Chemical and Biological Engineering, University of Colorado Boulder, Boulder, CO, 80309, United States}
\affil[6]{Open Molecular Software Foundation, Davis CA 95618, United States}
\affil[7]{Department of Bioengineering, University of California, Berkeley, Berkeley, CA, 94720, United States}
\affil[8]{Department of Pharmaceutical Sciences, University of California, Irvine, California 92697, United States}
\affil[9]{Tri-Institutional Ph.D. Program in Chemical Biology, Memorial Sloan Kettering Cancer Center, New York, New York 10065, United States}
\affil[10]{Simons Center for Computational Physical Chemistry and Center for Data Science, New York University, New York, N.Y. 10004, United States}

\corr{takaba.kb@om.asahi-kasei.co.jp}{KT}
\corr{john.chodera@choderalab.org}{JDC}
\corr{wangyq@wangyq.net}{YW}

\begin{document}

\maketitle

\begin{abstract}
The development of reliable and extensible molecular mechanics (MM) force fields---fast, empirical models characterizing the potential energy surface of molecular systems---is indispensable for biomolecular simulation and computer-aided drug design.
Here, we introduce a generalized and extensible machine-learned MM force field, \texttt{espaloma-0.3}, and an end-to-end differentiable framework using graph neural networks to overcome the limitations of traditional rule-based methods.
Trained in a single GPU-day to fit a large and diverse quantum chemical dataset of over 1.1M energy and force calculations, \texttt{espaloma-0.3} reproduces quantum chemical energetic properties of chemical domains highly relevant to drug discovery, including small molecules, peptides, and nucleic acids.
Moreover, this force field maintains the quantum chemical energy-minimized geometries of small molecules and preserves the condensed phase properties of peptides, self-consistently parametrizing proteins and ligands to produce stable simulations leading to highly accurate predictions of binding free energies.
This methodology demonstrates significant promise as a path forward for systematically building more accurate force fields that are easily extensible to new chemical domains of interest.

\end{abstract}
Molecular mechanics (MM) force fields~\cite{dauber2019biomolecular,hagler2019force} are fast, empirical models that describe the potential energy surfaces of biomolecular systems by treating them as collections of atomic point masses. 
These point masses interact via non-bonded and valence (bond, angle, and torsion) terms, which are typically parametrized to reproduce quantum chemical conformational energetics and physical properties. 
Despite their simplified representation of the underlying physical model, MM force fields have proven to be indispensable for a multitude of tasks in biomolecular simulation and computer-aided drug design~\cite{leach2001molecular,schlick2010molecular}, such as enumeration of putative bioactive conformations~\cite{coutsias2016exhaustive}, hit identification via virtual screening~\cite{bender2021practical}, prediction of membrane permeability~\cite{tse2019affordable}, simulations of biomolecular dynamics~\cite{prasad2018best}, and estimation of protein-ligand binding free energies via alchemical free energy calculations~\cite{mey2020best}.

\paragraph{Class~I MM force fields have been a widely popular compromise between speed and accuracy}
Class~I MM force fields~\cite{dauber2019biomolecular,hagler2019force} are most widely used for proteins, lipids, nucleic acids, and other relevant biomolecules due to the computational efficiency afforded by the simple functional form:
\begin{eqnarray}
\label{eq:u_mm}
    U_\text{MM}(\mathbf{x}; \Phi_\mathtt{FF}) 
    &= \sum\limits_{\text{bond}} & \frac{K_r}{2} (r_{i,j} - r_0)^2 \nonumber \\
    &+ \sum\limits_{\text{angle}} & \frac{K_\theta}{2} (\theta_{i,j,k} - \theta_0)^2 \nonumber \\
    &+ \sum\limits_{\text{torsion}} & \sum\limits_{n=1}^{n_\text{max}} K_{\phi, n}\left[1 + \cos(n \phi_{i,j,k,l} - \phi_0)\right] \nonumber \\
    &+ \sum\limits_{\text{Coulomb}} & \frac{1}{4 \pi \epsilon_0} \frac{q_i \, q_j}{r_{i,j}} \nonumber \\
    &+ \sum\limits_{\text{van der Waals}} & 4 \epsilon \left[ \left (\frac{\sigma_{i,j}}{r_{i,j}} \right)^{12} - \left(\frac{\sigma_{i,j}}{r_{i,j}}\right)^{6} \right],
\end{eqnarray}
where the total potential energy $U_\mathtt{MM}$ of a molecular system with coordinates $\mathbf{x}$ is defined by sets of force field parameters $\Phi_\mathtt{FF} = \{ K_r, K_\theta, r_0, \theta_0, K_{\phi, n}, \phi_0, q, \sigma, \epsilon\}_i$
specified for each atom $i$ or valence term (bond, angle, torsion) of the system. 
An out-of-plane term (an improper torsion) can be also introduced with the torsion term to improve molecular planarity. 
The van der Waals interactions are usually described with Lennard-Jones 12--6 potentials using the Lorentz-Berthelot~\cite{delhommelle2001inadequacy} combining rules to determine $\sigma$ and $\epsilon$ between different atom types, but alternative combination rules are possible. 
In practice, such interactions usually require combining distinct force field parameters developed independently for specific chemical domains to complement the heterogeneity of biomolecular systems.
Note that the functional forms of force fields can slightly differ among different Class~I force fields, incorporating different scaling constants and additional functional terms, such as CMAP~\cite{hagler2019force} and Urey-Bradley~\cite{dauber2019biomolecular}. 
The minimalistic nature of Class~I force fields has enabled them to achieve extraordinary speed on inexpensive hardware, with modern GPU-accelerated molecular simulation frameworks now able to generate more than 1 microsecond/day for many biomolecular drug targets~\cite{harvey2009acemd,salomon2013routine,eastman2017openmm} while still achieving useful accuracy in tasks such as predicting protein-ligand binding free energies for drug discovery~\cite{wang2015accurate,schindler2020large,gapsys2022pre}.

\paragraph{Traditional MM force field parametrization approaches struggle with complexity, limiting accuracy}

Traditionally, the construction of MM force fields requires expert knowledge of physical organic chemistry to build \textit{atom-typing} rules classifying atoms into discrete categories representing distinct chemical environments, enabling MM parameters to be subsequently assigned from a \textit{table} of relevant atomic, bond, angle, and torsion parameters. 
This creates an intractable mixed discrete-continuous optimization problem, posing a labor-intensive challenge, heavily reliant on human effort.
Force field accuracy is limited by the resolution of chemical perception, which in turn is limited by the number of distinct atom types. Attempting to improve accuracy by increasing the number of atom types results in a combinatorial explosion of bond, angle, and torsion parameters, which imposes strong practical limits~\cite{mobley2018escaping}.
As a result, modelers frequently turn to bespoke parameter generation tools---such as Paramfit~\cite{betz2015paramfit}, FFBuilder~\cite{harder2016opls3} or OpenFF BespokeFit~\cite{bespoke}---to assign individual parameters for molecules of interest for which high accuracy is needed, requiring expensive quantum chemical calculations to be performed \textit{ad hoc} and diminishing the speed benefits of Class~I force fields.

\paragraph{Traditional MM force field parametrization approaches often aim for divide-and-conquer, rather than self-consistency}
To tame the explosion of atom type complexity, biomolecular force field efforts have frequently taken the approach of building separate but purportedly compatible models for proteins, small molecules, and other biomolecules independently.
For example, the recent AmberTools~23 release~\cite{amber2023} recommends combining independently developed force fields to simulate systems containing proteins~\cite{ff14SB}, DNA~\cite{OL15,OL15full}, RNA~\cite{OL3}, water~\cite{jorgensen1983comparison,TIP4P-EW,OPC}, monovalent~\cite{joung2008determination,joung2009molecular} and divalent~\cite{li2013rational,li2014taking,li2015parameterization} counterions, lipids~\cite{gould2018lipid17}, carbohydrates~\cite{kirschner2008glycam06}, glycoconjugates~\cite{demarco2009atomic,demarco2010presentation}, small molecules~\cite{wang2004development,wang2006automatic}, post-translational modifications~\cite{khoury2013forcefield_ptm}, and nucleic acid modifications~\cite{aduri2007rna}---which collectively might represent more than 100 person-years of effort.
While this simplifies the subproblems of parametrizing each class of molecules, using these separate force fields together to treat complex, heterogeneous systems is neither simple nor optimal.
There are often overlaps in the chemical space that each force field is designed to model, with no guarantee that the parameters in these regions are identical and remain entirely compatible. 
This introduces significant caveats when multiple classes of biomolecules interact, risking poor accuracy and greatly frustrating the cases where molecules of different classes must be covalently bonded. 
As such, extension or expansion to new classes of biomolecules or chemical spaces becomes a time-consuming ordeal, as combining force fields often results in a large combinatorial space of possible force field parameters where the quality of the resulting force field depends heavily on the choices made by the user.


There have been numerous efforts to systematize and automate the process of force field development~\cite{mobley2018escaping,wang2014building,amberfb15,parsley,sage,harder2016opls3}. 
For example, the Open Force Field Initiative has developed a number of modern, open-source tools~\cite{bespoke,openffevaluator}, datasets, and force fields~\cite{parsley,sage} that employ a direct approach to chemical perception~\cite{mobley2018escaping}, which use a standard SMARTS-based chemical substructure query to assign entire sets of valence parameters (atoms, bonds, angles, torsions) in a hierarchical manner, attempting to ameliorate the combinatorial explosion of parameters.
There have also been extensive efforts to systematically optimize parameters using finite-difference methods~\cite{wang2014building,amberfb15} and machine learning approaches~\cite{befort2021machine,wang2023dmff}. 
However, much of the work focuses on small molecules, and extending the force field to new chemical domains still requires human effort---jointly optimizing discrete chemical perception rules and continuous force field parameters remains intractable.

\paragraph{A graph neural network parametrization scheme can automate, simplify, and significantly improve the accuracy of MM force fields with no performance penalty}
Recently, we proposed a novel approach---Espaloma~\cite{D2SC02739A} (\textit{\textbf{e}xtensible \textbf{s}urrogate \textbf{p}otenti\textbf{al} \textbf{o}ptimized by \textbf{m}es-s\textbf{a}ge passing})---which replaces the rule-based discrete \textit{atom-typing} schemes with a continuous atomic representation generated by graph neural networks that operate on chemical graphs~\cite{D2SC02739A,thurlemann2023regularized, wangthesis}. 
These atom representations are coupled with a set of symmetry-preserving pooling layers and feed-forward neural networks to enable fully end-to-end differentiable construction of MM force fields. 
The neural network parameters are optimized directly using standard machine learning frameworks to fit quantum chemical and/or experimental data. 
The expressiveness of Espaloma's continuous atomic representations eliminates the need to combine force fields developed for different chemical domains. 
Thus, Espaloma can \textit{self-consistently} parametrize any system of molecules with elemental coverage in its training set.

Earlier work~\cite{D2SC02739A,thurlemann2023regularized} demonstrated that this approach, in principle, parametrizes multiple classes of biomolecules---the open source Espaloma package was used to train a small Espaloma model for a Class~I MM force field on a limited set of 45\,000 quantum chemical calculations covering small molecules and amino acids~\cite{D2SC02739A}. 
While surprisingly robust in comparison to traditional small molecule and amino acid force fields, that model was far from providing comprehensive coverage of chemical space relevant to biomolecular modeling and drug discovery, and its potential usage for real-world applications remained unclear. 

\paragraph{espaloma-0.3: a versatile, robust, and accurate machine-learned Class~I MM force field retrainable in a single-GPU day}
In this paper, we introduce a significantly enhanced Espaloma framework that incorporates energy and force matching with quantum chemical data, scalability to massive quantum chemical datasets, and stringent regularization for enhanced model stability.
We demonstrate how this approach can easily fine-tune an existing Class~I small molecule force field and extend to new chemical domains of interest without a performance penalty, resulting in a generalized and extensible machine-learned Class~I MM force field, \texttt{espaloma-0.3}.
Trained in a single GPU-day to fit a large and diverse curated quantum chemical dataset of over 1.1M energy and force calculations for 17\,000 unique molecular species, \texttt{espaloma-0.3} reproduces quantum chemical energetic properties of chemical spaces of small molecules, peptides, and nucleic acids much more accurately than the well-established MM force fields widely used in the fields of biomolecular simulation and computer-aided drug design.
Furthermore, it maintains the quantum chemical energy-minimized geometries of small molecules and preserves the condensed phase properties of peptides, thus self-consistently parametrizing proteins and ligands to produce stable simulations leading to highly accurate protein-ligand binding free energy predictions.
To our knowledge, this study represents the \textit{first} well-demonstrated example of a self-consistent MM force field capable of parametrizing a protein-ligand system that is applicable for real-world drug discovery purposes.

This enhanced Espaloma framework demonstrates significant promise as a path forward for systematically building more accurate and extensible force fields with additional quantum chemical data, similarly to how foundational large language models can be fine-tuned to perform better on domain tasks of interest.

\section{Espaloma provides a flexible, end-to-end differentiable framework for assigning molecular mechanics (MM) parameters using graph neural networks (GNNs)}

\begin{figure}[tb]
    \centering
    \includegraphics[width=0.9\textwidth]{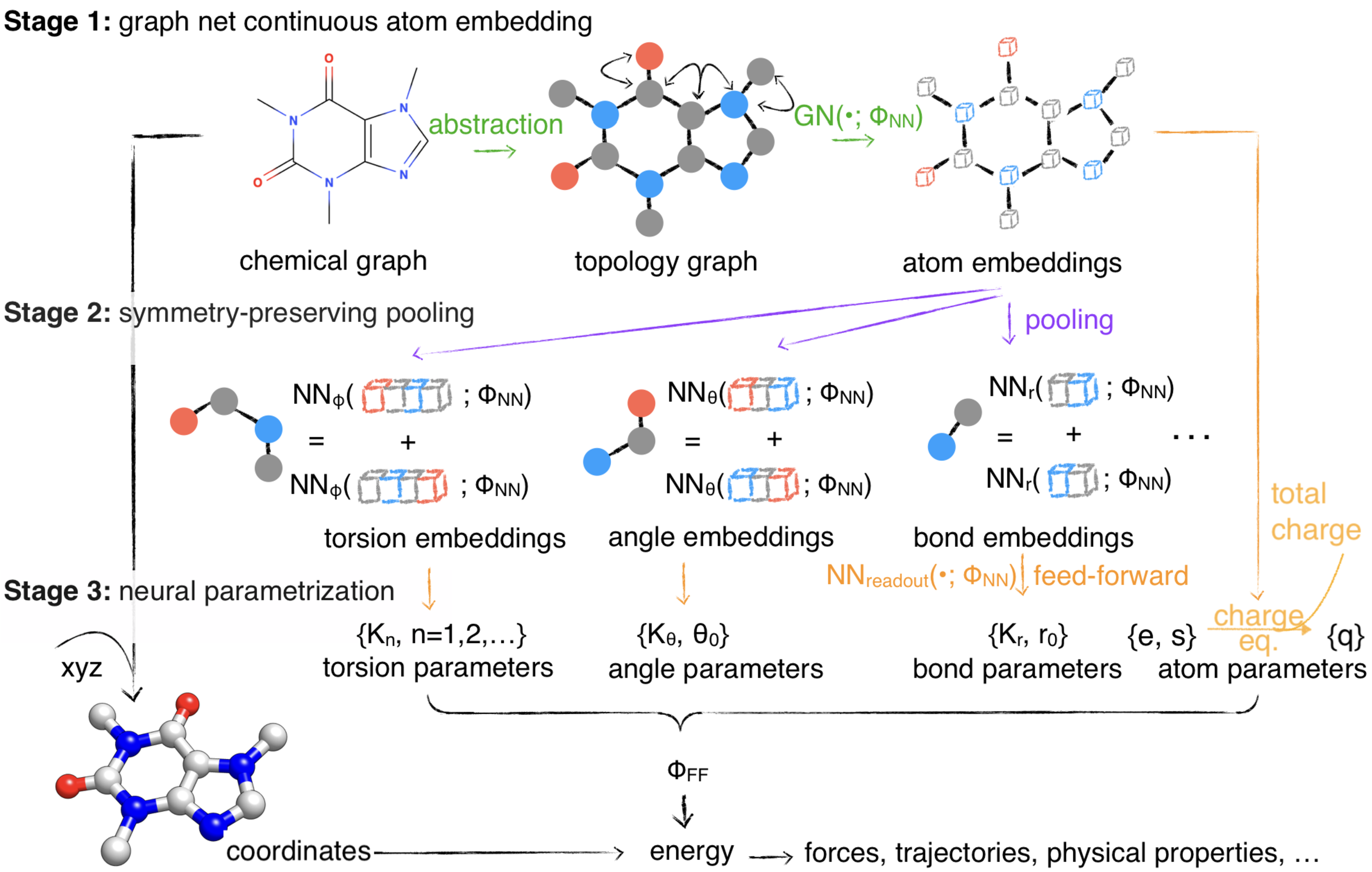}
    \caption{\label{fig:abstract}
    \textbf{Espaloma is an end-to-end differentiable molecular mechanics parameter assignment scheme for arbitrary organic molecules.} 
    Espaloma (\textit{\textbf{e}xtensible \textbf{s}urrogate \textbf{p}otenti\textbf{al} \textbf{o}ptimized by \textbf{m}ess\textbf{a}ge-passing}) is a modular approach for directly computing molecular mechanics force field parameters $\Phi_\mathrm{FF}$ from a chemical graph $\mathcal{G}$ such as a small molecule or biopolymer via a process that is fully differentiable in the model parameters $\Phi_\mathrm{NN}$.
    In {\bf Stage 1}, a graph neural network is used to generate continuous latent atom embeddings describing local chemical environments from the chemical graph.
    In {\bf Stage 2}, these atom embeddings are transformed into feature vectors that preserve appropriate symmetries for atom, bond, angle, and proper/improper torsion inference via Janossy pooling~\cite{murphy2018janossy}.
    In {\bf Stage 3}, molecular mechanics parameters are directly predicted from these feature vectors using feed-forward neural networks.
    This parameter assignment process is performed once per molecular species, allowing the potential energy to be rapidly computed using standard molecular mechanics or molecular dynamics frameworks thereafter.
    The collection of parameters $\Phi_\mathrm{NN}$ describing the espaloma model can be considered as the equivalent complete specification of a traditional molecular mechanics force field such as GAFF~\cite{wang2004development,wang2006automatic}/AM1-BCC~\cite{jakalian2000fast,jakalian2002fast} in that it encodes the equivalent of traditional typing rules, parameter assignment tables, and even partial charge models.
    Figure reproduced from the arXiv preprint of \citet{D2SC02739A} under the \href{http://arxiv.org/licenses/nonexclusive-distrib/1.0/}{arXiv non-exclusive license}.
    }
\end{figure}

Espaloma~\cite{D2SC02739A} ({\bf Figure~\ref{fig:abstract}}) operates analogously to an \textit{atom-typing} based force field, where chemical perceptions are predefined to generate MM force field parameters $\Phi_\text{FF}$. However, instead of working with atom types, Espaloma operates on a chemical graph $\mathcal{G}$ using a graph neural network (GNN) parametrized by neural network model parameters $\Phi_\text{NN}$,
\begin{equation}
\Phi_\text{FF} \leftarrow \texttt{espaloma} (\mathcal{G}, \Phi_\text{NN}) .
\end{equation}
The resulting parameters $\Phi_\text{FF}$ can then be subsequently used in a standard molecular mechanics package to compute the MM energy and forces for any conformation, as with a standard MM force field.

Espaloma parametrizes molecular systems in three sequential stages ({\bf Figure~\ref{fig:abstract}}):

\textbf{Stage 1: Graph neural networks generate a continuous vectorial atom embedding, replacing discrete atom-typing rules.}
First, using chemoinformatics toolkits such as RDKit~\cite{rdkit}, the molecular system is abstracted as a \textit{graph}, with nodes and edges denoted as atoms and covalent bonds, respectively.
Espaloma uses GNNs~\cite{duvenaud2015convolutional, kipf2016gcn, gilmer2017neural, xu2018powerful, battaglia2018relational, jain2018topology, wu2019simplifying, wang2019deep, wang2019dynamic, joshi2023expressive} as a replacement for rule-based chemical environment perception~\cite{mobley2018escaping} to operate on this graph.
These neural architectures learn useful representations of atomic chemical environments from simple input features by updating and pooling embedding vectors via message-passing iterations to neighboring atoms ~\cite{gilmer2017neural}.
As such, symmetries in chemical graphs (chemical equivalencies) are inherently preserved, while a rich, continuous, and differentiably learnable representation of the atomic environment is derived.

\textbf{Stage 2: Symmetry-preserving pooling generates continuous bond, angle, and torsion embeddings.}
The representations determined by GNNs in {\bf Stage 1} are used to come up with bond, angle, and torsion representations in a symmetry-preserving manner, where the relevant equivalent atom permutations are listed and summed up via Janossy pooling~\cite{murphy2018janossy}.

\textbf{Stage 3: Neural parametrization of atoms, bonds, angles, and torsions replaces tabulated parameter assignment.}
In the final stage, feed-forward neural networks learn the mapping from these symmetry-preserving invariant atom, bond, angle, and torsion embeddings to MM parameters $\Phi_\text{FF}$ that reflect the specific chemical environments appropriate for these terms.
Each distinct parameter class (such as atom, bond, angle, and torsion parameters) is assigned by a separate neural network, making this stage fully modular.
This stage is analogous to the final table lookup step in traditional force field construction, but it offers significant added flexibility due to the continuous embedding that captures the chemical environment specific to the assigned potential energy term.

The final output is a set of force field parameters $\Phi_\mathtt{FF}$ uniquely determined by the neural network conditioned on its associated weights $\Phi_\mathtt{NN}$. This means that once 
the $\Phi_\mathtt{NN}$ is optimized, biomolecular simulations can be performed as fast as those simulated with traditional MM force fields.
Atomic partial charges can also be generated within the Espaloma framework, using a geometry-independent charge equilibration approach~\cite{gilson2003fast} to rapidly generate AM1-BCC~\cite{jakalian2000fast,jakalian2002fast} quality charges~\cite{wang2019graph, wang2023espalomacharge}.

Overall, the Espaloma framework is end-to-end differentiable---the error in energy (or the function thereof, such as forces) can be backpropagated to optimize the force field parameters $\Phi_\mathtt{FF}$, and thereby neural network parameters $\Phi_\mathtt{NN}$ that govern how they are produced from the input molecule. {\bf Stage 3} is especially modular and flexible. New force field terms that act on atoms, bonds, angles, torsions, or combinations thereof can easily be added and the entire force field refit starting from either an existing $\Phi_\mathtt{NN}$ or training from scratch.
In this way, Espaloma provides a rapid and flexible approach to experimenting with different potential functions (such as the addition of point polarizability or exploration of alternative functional forms) or retraining with augmented training datasets.

\section{Extensive open quantum chemical dataset curated to provide coverage of biomolecules: small molecules, proteins, and nucleic acids}
\label{sec:dataset}
To develop a self-consistent MM force field broadly applicable to biomolecular modeling, we first curate a high-quality gas-phase quantum chemical dataset deposited in QCArchive~\cite{qcarchive} ({\bf Table~\ref{tab:rmse}}). 
The curated quantum chemical dataset is built from several components that provide complementary coverage of relevant biomolecular chemistries: 
from the \textit{foundational} SPICE dataset~\cite{eastman2023spice}, we extracted a large set of drug-like small molecules selected from PubChem~\cite{li2010pubchem}, dipeptides (capped 2-mers) and their common protonation and tautomeric variants, and diverse molecular fragments providing broad coverage of biomolecules from the DES370K dataset~\cite{DES370K};
from the OpenFF 1.x ("Parsley")~\cite{parsley} and 2.x ("Sage")~\cite{sage} datasets, we extracted optimization and torsion-drive datasets for diverse small molecules;
a diverse set of dipeptide (capped 2-mers), tripeptides (capped 3-mers), disulfide-bridged, bioactive, and cyclic peptides from the PepConf dataset~\cite{prasad2019pepconf};
a peptide torsion scan set generated by the Open Force Field Consortium for the OpenFF 3.x ("Rosemary") force field~\cite{rosemary};
and a new set of RNA nucleosides, trinucleotides, and diverse experimental RNA fragments sourced from the Nucleic Acid Database~\cite{nucleic-acid-database} and RNA Structure Atlas~\cite{rna-structure-atlas} to extend coverage to this important and growing class of drug targets. 

To capture the rugged conformational energy surface of biomolecules, the quantum chemical datasets were extracted from three different QCArchive workflows: \texttt{Dataset}, \texttt{OptimizationDataset}, and \texttt{TorsionDriveDataset}. 
A \texttt{Dataset} contains single-point energy calculations of structures that are not necessarily at their local quantum energy minima, generated using MD simulations or conformer generators. 
An \texttt{OptimizationDataset} is a collection of QM optimization trajectories for a given structure. 
A \texttt{TorsionDriveDataset} involves torsion scans performed on a set of rotatable torsions, followed by QM optimization.

The curated dataset consists of 1\,188\,317 conformations of 17\,427 unique molecules in total. 
We also computed the AM1-BCC ELF10 partial charges using the OpenEye Toolkits to train and generate AM1-BCC~\cite{jakalian2000fast,jakalian2002fast} quality partial charges with Espaloma.
Complete details of the dataset construction and composition are given in {\bf SI Section~\ref{sec:si_dataset}}. 
All quantum chemical energies are computed with the Open Force Field (OpenFF) standard level of quantum chemical theory (B3LYP-D3BJ/DZVP)~\cite{parsley,sage}, which balances the computational efficiency and accuracy to reproduce the conformations generated by higher levels of theories~\cite{pavan2022openffbench}. 
These quantum chemical datasets were generated with the open source \texttt{psi4} quantum chemistry package~\cite{smith2020psi4} using the QCArchive~\cite{qcarchive} QCFractal infrastructure via OpenFF QCSubmit~\cite{qcsubmit} workflows.

\begin{table}[tbp]
\centering
\resizebox{\columnwidth}{!}{%
\begin{tabular}{ccccccccccc}
\toprule
  {\begin{tabular}[c]{c@{}}
      Dataset\\
      {\small (QCArchive Workflow)}
  \end{tabular}} &  
  Category &
  Mols &
  Confs &
  Split &
  \multicolumn{2}{c}
    {\begin{tabular}[c]{@{}c@{}}
        {\bf \large espaloma-0.3}\\ 
        {\footnotesize Energy RMSE (kcal/mol)}\\ 
        {\footnotesize Force RMSE (kcal/mol $\cdot$ ${\text{Å}}^{-1}$)}
    \end{tabular}} &
  \multicolumn{4}{c}
    {\begin{tabular}[c]{@{}c@{}}{\bf \large Baseline Force Field} {\large (Test molecules)}\\ 
    {\footnotesize Energy RMSE (kcal/mol)}\\ 
    {\footnotesize Force RMSE (kcal/mol $\cdot$ ${\text{Å}}^{-1}$)}
    \end{tabular}} \\
  \cmidrule(r){6-7} \cmidrule(r){8-11}
   &
   &
   &
   &
   &
  Train (80\%) &
  Test (10\%) &
  gaff-2.11~\cite{gaff2} &
  openff-2.0.0~\cite{openff-2.0.0} &
  openff-2.1.0~\cite{openff-2.1.0} &
  ff14SB~\cite{ff14SB}/RNA.OL3~\cite{OL3} \\
  \midrule
  \begin{tabular}[c]{@{}c@{}}{\bf SPICE-Pubchem~\cite{eastman2023spice, kim2023pubchem}}\\{\small (Dataset)}\end{tabular} &
  \begin{tabular}[c]{@{}c@{}}Small\\ molecule\end{tabular} &
  14110 &
  608436 &
  {\footnotesize 80:10:10} &
  \begin{tabular}[c]{@{}c@{}}
    $2.06_{2.04}^{2.07}$\\$6.22_{6.19}^{6.26}$
  \end{tabular} &
  \begin{tabular}[c]{@{}c@{}}
    $2.30_{2.25}^{2.36}$\\$6.81_{6.68}^{6.95}$
  \end{tabular} &
  \begin{tabular}[c]{@{}c@{}}
    $4.39_{4.30}^{4.48}$\\$14.02_{13.71}^{14.37}$
  \end{tabular} &
  \begin{tabular}[c]{@{}c@{}}
    $4.21_{4.13}^{4.30}$\\$13.95_{13.71}^{14.20}$
  \end{tabular} &
  \begin{tabular}[c]{@{}c@{}}
    $4.45_{4.37}^{4.53}$\\$15.45_{15.17}^{15.75}$
  \end{tabular} &
  \begin{tabular}[c]{@{}c@{}}
    ---\\---
  \end{tabular} 
  \vspace{1mm} \\

  \begin{tabular}[c]{@{}c@{}}{\bf SPICE-DES-Monomers~\cite{eastman2023spice, DES370K}}\\{\small (Dataset)}\end{tabular} &
  \begin{tabular}[c]{@{}c@{}}Small\\ molecule\end{tabular} &
  369 &
  18435 &
  {\footnotesize 80:10:10} &
  \begin{tabular}[c]{@{}c@{}}
    $1.39_{1.32}^{1.46}$\\$5.86_{5.69}^{6.02}$
  \end{tabular} &
  \begin{tabular}[c]{@{}c@{}}
    $1.36_{1.13}^{1.67}$\\$5.91_{5.49}^{6.42}$
  \end{tabular} &
  \begin{tabular}[c]{@{}c@{}}
    $1.88_{1.57}^{2.22}$\\$9.46_{8.09}^{10.91}$
  \end{tabular} &
  \begin{tabular}[c]{@{}c@{}}
    $2.34_{1.97}^{2.75}$\\$11.12_{9.86}^{12.47}$
  \end{tabular} &
  \begin{tabular}[c]{@{}c@{}}
    $2.43_{2.05}^{2.81}$\\$11.87_{10.57}^{13.15}$
  \end{tabular} &
  \begin{tabular}[c]{@{}c@{}}
    ---\\---
  \end{tabular} 
  \vspace{1mm} \\

  \begin{tabular}[c]{@{}c@{}}{\bf Gen2-Opt}\\{\small 
  (OptimizationDataset)}\end{tabular} &
  \begin{tabular}[c]{@{}c@{}}Small\\ molecule\end{tabular} &
  1024 &
  244989 &
  {\footnotesize 80:10:10} &
  \begin{tabular}[c]{@{}c@{}}
    $1.36_{1.26}^{1.48}$\\$3.94_{3.79}^{4.11}$
  \end{tabular} &
  \begin{tabular}[c]{@{}c@{}}
    $1.66_{1.21}^{2.29}$\\$4.47_{3.90}^{5.40}$
  \end{tabular} &
  \begin{tabular}[c]{@{}c@{}}
    $2.29_{1.88}^{2.82}$\\$10.51_{9.75}^{11.36}$
  \end{tabular} &
  \begin{tabular}[c]{@{}c@{}}
    $2.18_{1.73}^{2.77}$\\$10.53_{9.86}^{11.40}$
  \end{tabular} &
  \begin{tabular}[c]{@{}c@{}}
    $2.25_{1.78}^{2.85}$\\$11.67_{10.83}^{12.53}$
  \end{tabular} &
  \begin{tabular}[c]{@{}c@{}}
    ---\\---
  \end{tabular} 
  \vspace{1mm} \\

  \begin{tabular}[c]{@{}c@{}}{\bf Gen2-Torsion}\\{\small 
  (TorsionDriveDataset)}\end{tabular} &
  \begin{tabular}[c]{@{}c@{}}Small\\ molecule\end{tabular} &
  729 &
  25832 &
  {\footnotesize 80:10:10} &
  \begin{tabular}[c]{@{}c@{}}
    $1.76_{1.61}^{1.91}$\\$4.31_{4.18}^{4.44}$
  \end{tabular} &
  \begin{tabular}[c]{@{}c@{}}
    $1.64_{1.32}^{2.01}$\\$4.71_{4.18}^{5.29}$
  \end{tabular} &
  \begin{tabular}[c]{@{}c@{}}
    $2.53_{1.95}^{3.21}$\\$10.50_{9.42}^{11.67}$
  \end{tabular} &
  \begin{tabular}[c]{@{}c@{}}
    $1.69_{1.38}^{2.06}$\\$11.11_{10.21}^{12.09}$
  \end{tabular} &
  \begin{tabular}[c]{@{}c@{}}
    $1.83_{1.46}^{2.24}$\\$11.92_{11.04}^{12.87}$
  \end{tabular} &
  \begin{tabular}[c]{@{}c@{}}
    ---\\---
  \end{tabular} 
  \vspace{1mm} \\

  \begin{tabular}[c]{@{}c@{}}{\bf SPICE-Dipeptide~\cite{eastman2023spice}}\\{\small (Dataset)}\end{tabular} &
  Peptide &
  677 &
  26279 &
  {\footnotesize 80:10:10} &
  \begin{tabular}[c]{@{}c@{}}
    $3.21_{3.16}^{3.26}$\\$7.98_{7.88}^{8.07}$
  \end{tabular} &
  \begin{tabular}[c]{@{}c@{}}
    $3.09_{2.96}^{3.21}$\\$7.78_{7.55}^{8.02}$
  \end{tabular} &
  \begin{tabular}[c]{@{}c@{}}
    $4.24_{4.07}^{4.42}$\\$11.90_{11.50}^{12.32}$
  \end{tabular} &
  \begin{tabular}[c]{@{}c@{}}
    $4.11_{3.96}^{4.28}$\\$11.95_{11.62}^{12.32}$
  \end{tabular} &
  \begin{tabular}[c]{@{}c@{}}
    $4.28_{4.10}^{4.44}$\\$11.57_{11.26}^{11.88}$
  \end{tabular} &
  \begin{tabular}[c]{@{}c@{}}
    $4.36_{4.20}^{4.55}$\\$11.76_{11.40}^{12.09}$
  \end{tabular} 
  \vspace{1mm} \\

  \begin{tabular}[c]{@{}c@{}}{\bf Pepconf-Opt~\cite{prasad2019pepconf}}\\{\small 
  (OptimizationDataset)}\end{tabular} &
  Peptide &
  557 &
  166291 &
  {\footnotesize 80:10:10} &
  \begin{tabular}[c]{@{}c@{}}
    $2.61_{2.43}^{2.83}$\\$3.83_{3.60}^{4.09}$
  \end{tabular} &
  \begin{tabular}[c]{@{}c@{}}
    $2.79_{2.45}^{3.13}$\\$4.01_{3.63}^{4.46}$
  \end{tabular} &
  \begin{tabular}[c]{@{}c@{}}
    $3.53_{3.03}^{3.82}$\\$8.07_{7.84}^{8.23}$
  \end{tabular} &
  \begin{tabular}[c]{@{}c@{}}
    $2.91_{2.56}^{3.39}$\\$8.74_{8.49}^{9.08}$
  \end{tabular} &
  \begin{tabular}[c]{@{}c@{}}
    $3.19_{2.73}^{3.66}$\\$8.79_{8.27}^{9.56}$
  \end{tabular} &
  \begin{tabular}[c]{@{}c@{}}
    $3.59_{3.00}^{4.17}$\\$9.13_{8.67}^{9.70}$
  \end{tabular} 
  \vspace{1mm} \\

  \begin{tabular}[c]{@{}c@{}}{\bf Protein-Torsion}\\{\small (TorsionDriveDataset)}\end{tabular} &
  Peptide &
  62 &
  48999 &
  {\footnotesize 80:10:10} &
  \begin{tabular}[c]{@{}c@{}}
    $2.27_{2.06}^{2.50}$\\$3.94_{3.70}^{4.24}$
  \end{tabular} &
  \begin{tabular}[c]{@{}c@{}}
    $1.93_{1.73}^{2.14}$\\$3.49_{3.22}^{3.78}$
  \end{tabular} &
  \begin{tabular}[c]{@{}c@{}}
    $3.53_{3.03}^{3.82}$\\$8.07_{7.84}^{8.23}$
  \end{tabular} &
  \begin{tabular}[c]{@{}c@{}}
    $2.91_{2.56}^{3.39}$\\$8.74_{8.49}^{9.00}$
  \end{tabular} &
  \begin{tabular}[c]{@{}c@{}}
    $3.19_{2.73}^{3.66}$\\$8.79_{8.27}^{9.56}$
  \end{tabular} &
  \begin{tabular}[c]{@{}c@{}}
    $3.59_{3.00}^{4.17}$\\$9.13_{8.67}^{9.70}$
  \end{tabular} 
  \vspace{1mm} \\

  \begin{tabular}[c]{@{}c@{}}{\bf RNA-Diverse}\\{\small (Dataset)}\end{tabular} &
  RNA &
  64 &
  3703 &
  {\footnotesize 80:10:10} &
  \begin{tabular}[c]{@{}c@{}}
    $4.12_{3.95}^{4.31}$\\$4.44_{4.40}^{4.47}$
  \end{tabular} &
  \begin{tabular}[c]{@{}c@{}}
    $4.17_{3.85}^{4.52}$\\$4.41_{4.29}^{4.51}$
  \end{tabular} &
  \begin{tabular}[c]{@{}c@{}}
    $5.65_{4.95}^{6.32}$\\$17.19_{16.71}^{17.71}$
  \end{tabular} &
  \begin{tabular}[c]{@{}c@{}}
    $5.79_{5.37}^{6.19}$\\$18.54_{17.85}^{19.10}$
  \end{tabular} &
  \begin{tabular}[c]{@{}c@{}}
    $6.26_{5.64}^{6.90}$\\$19.68_{19.19}^{20.15}$
  \end{tabular} &
  \begin{tabular}[c]{@{}c@{}}
    $6.06_{5.70}^{6.43}$\\$19.38_{18.77}^{19.83}$
  \end{tabular} 
  \vspace{1mm} \\

  \begin{tabular}[c]{@{}c@{}}{\bf RNA-Trinucleotide}\\{\small (Dataset)}\end{tabular} &
  RNA &
  64 &
  35811 &
  {\footnotesize 0:0:100} &
  \begin{tabular}[c]{@{}c@{}}
    ---\\---
  \end{tabular} &
  \begin{tabular}[c]{@{}c@{}}
    $3.75_{3.59}^{3.94}$\\$4.28_{4.20}^{4.39}$
  \end{tabular} &
  \begin{tabular}[c]{@{}c@{}}
    $5.79_{5.61}^{5.98}$\\$17.15_{17.00}^{17.28}$
  \end{tabular} &
  \begin{tabular}[c]{@{}c@{}}
    $5.81_{5.67}^{5.96}$\\$18.88_{18.72}^{19.02}$
  \end{tabular} &
  \begin{tabular}[c]{@{}c@{}}
    $6.26_{6.10}^{6.42}$\\$19.97_{19.81}^{20.13}$
  \end{tabular} &
  \begin{tabular}[c]{@{}c@{}}
    $5.94_{5.77}^{6.12}$\\$19.82_{19.67}^{19.97}$
  \end{tabular} 
  \vspace{1mm} \\

  \begin{tabular}[c]{@{}c@{}}{\bf RNA-Nucleoside}\\{\small (Dataset)}\end{tabular} &
  RNA &
  4 &
  9542 &
  {\footnotesize 100:0:0} &
  \begin{tabular}[c]{@{}c@{}}
    $1.32_{1.16}^{1.49}$\\$4.17_{3.86}^{4.47}$
  \end{tabular} &
  \begin{tabular}[c]{@{}c@{}}
    ---\\---
  \end{tabular} &
  \begin{tabular}[c]{@{}c@{}}
    ---\\---
  \end{tabular} &
  \begin{tabular}[c]{@{}c@{}}
    ---\\---
  \end{tabular} &
  \begin{tabular}[c]{@{}c@{}}
    ---\\---
  \end{tabular} &
  \begin{tabular}[c]{@{}c@{}}
    ---\\---
  \end{tabular} \\
\bottomrule
\end{tabular}%
}
\caption{{\bf \texttt{espaloma-0.3} can directly fit quantum chemical potential energies and forces more accurately than baseline force fields.}
Espaloma was fit to quantum chemical (QC) potential energies and forces from various gas-phase QC datasets sourced from QCArchive~\cite{qcarchive}, covering a broad chemical space that includes small molecules, peptides, and RNA molecules (see {\bf SI Section~\ref{sec:si_dataset}}). 
The entire dataset consists of 17\,427 unique molecules and 1\,188\,317 conformations. 
These datasets were extracted from three different QCArchive workflows: \texttt{BasicDataset}, \texttt{OptimizationDataset}, and \texttt{TorsionDriveDataset}. The datasets were partitioned into train, validate, and test sets in an 80:10:10 ratio split by molecules, except for the \texttt{RNA-Trinucleotide} and \texttt{RNA-Nucleoside} datasets. 
Since RNA nucleosides and trinucleosides lack chemical diversity, the \texttt{RNA-Nucleoside} dataset was used for training, whereas the \texttt{RNA-Trinucleotide} dataset, which covers the same molecules as the \texttt{RNA-Diverse} dataset but with much more diverse conformers, was used as a test set. The number of molecules and total conformations for each dataset is annotated in the table. We report the root mean square error (RMSE) on the training and test sets, along with the performance of other force fields as baselines on the test set. The baseline force fields used were \texttt{gaff-2.11}~\cite{gaff2}, \texttt{openff-2.0.0}~\cite{openff-2.0.0}, and \texttt{openff-2.1.0}~\cite{openff-2.1.0} for small molecules, Amber \texttt{ff14SB}~\cite{ff14SB} for peptides, and Amber \texttt{RNA.OL3}~\cite{OL3} for RNA molecules. All statistics are computed with predicted and reference energies centered to have a zero mean for each molecule similar to the previous work~\cite{D2SC02739A}. The 95\% confidence intervals, as annotated in the results, were calculated by bootstrapping molecule replacement using 1000 replicates.
\vspace{-20 pt}
}
\label{tab:rmse}
\end{table}

\section{Espaloma force field reproduces quantum chemical energies and forces}
\label{sec:rmse}

%
%
Leveraging the curated gas-phase quantum chemical datasets discussed in {\bf Section~\ref{sec:dataset}}, we fine-tune and extend the OpenFF 2.0 ("Sage") force field, \texttt{openff-2.0.0}---a Class~I MM force field originally developed for small molecules---into new chemical domains of interest, resulting in a novel Class~I MM force field termed \texttt{espaloma-0.3}.
Similar to the original implementation~\cite{D2SC02739A} and historic practice in MM force field parametrization~\cite{parsley,amberfb15,ff14SB,ff19SB,OL15full,OL3}, we optimized the valence parameters (bonds, angles, and proper/improper torsions) and use the Lennard-Jones parameters from \texttt{openff-2.0.0}~\cite{sage}. 
While it is possible to optimize Lennard-Jones parameters as well, it is critical to include more computationally expensive condensed-phase simulations when doing so~\cite{boulanger2018optimized,boothroyd2022improving}.
For partial charges, following the protocol of \citet{wang2019graph}, we predict the electronegativity and hardness of atoms used in a \textit{charge equilibration}~\cite{gilson2003fast} to predict atomic partial charges while preserving the total charge of a given molecule. 
We utilize the AM1-BCC ELF10 partial charges computed with the OpenEye Toolkits as our target partial charges.

We enhance the original Espaloma framework to improve the model stability and data efficiency (see {\bf SI Section~\ref{sec:si_refit}} for further details):
\begin{itemize}
    \item quantum chemical forces are incorporated into training to provide more information about the quantum chemical potential surface;
    \item \textit{L2} regularization is applied to proper and improper torsion force constants to suppress spurious features in torsion profiles; 
    \item improper torsion terms expressed using $n=1, 2$ periodicities to reduce the complexity of the model and to align with other conventional force fields which usually employs $n=1, 2$ periodicities;
    \item node features that were sensitive to resonance form have been eliminated to ensure chemically equivalent representations of the same molecule receive identical parameters.
\end{itemize}
To train \texttt{espaloma-0.3}, we randomly shuffle the datasets and split each dataset by molecules into train, test, and validation sets (80\%, 10\%, and 10\%, respectively) based on unique isomeric SMILES strings.
Since the MM force field is incapable of reproducing quantum chemical heats of formation, which are reflected as an additive offset in quantum chemical energy targets for each molecule, we shift the reference quantum chemical energy of each molecule to have zero mean;
note that when deployed, the absolute value of MM energy is not physically meaningful and traditional MM force fields are never used to simulate bond-breaking events.
The loss function used in training included deviations from quantum chemical snapshot energies and forces, as well as deviations from target partial charges for each molecule in the training set (see see {\bf SI Section~\ref{sec:si_refit}} for complete details).

%
%


As shown in {\bf Table~\ref{tab:rmse}}, \texttt{espaloma-0.3} significantly outperforms all baseline force fields (\texttt{gaff-2.11}~\cite{gaff2}, \texttt{openff-2.0.0}~\cite{openff-2.0.0}, \texttt{openff-2.1.0}~\cite{openff-2.1.0}, Amber \texttt{ff14SB}~\cite{ff14SB}, Amber ~\texttt{RNA.OL3}~\cite{OL3}) in reproducing quantum chemical energies and forces, demonstrating the ability of \texttt{espaloma-0.3} to recapitulate the quantum chemical energy surface more accurately than current-generation Class~I MM potentials for biomolecules and organic chemistry despite using the same functional form.
In contrast, the baseline force fields widely popular in the field of biomolecular simulations yield considerable energy errors and huge force errors (on average twice to thrice that of \texttt{espaloma-0.3}) with respect to quantum chemical calculations.
The performance superiority holds true across diverse chemical categories, suggesting the general utility of \texttt{espaloma-0.3} in a wide array of chemical and biochemical modeling tasks, as evidenced in {\bf Section~\ref{sec:binding}} and {\bf Section~\ref{sec:vanilla}}.
These observations hold true when Espaloma is trained with different data splitting strategies ({\bf SI~Table~\ref{tab:si_rmse}}).

Notably, the backbone torsion parameters for \texttt{ff14SB} are empirically adjusted to improve agreement with condensed-phase NMR data. Therefore, it might be expected to perform less effectively when benchmarked against quantum chemical energetic properties. For a more rigorous comparison, we conducted the same benchmark experiment using \texttt{ff14SBonlysc}~\cite{ff14SBonlysc}, which is the same model as \texttt{ff14SB} but without the empirical backbone corrections. The resulting energy RMSE on test datasets for \texttt{SPICE-Dipeptide}, \texttt{Pepconf-Opt}, and \texttt{Protein-Torsion} were 4.36 [95\% CI: 4.52, 4.19], 3.93 [95\% CI: 3.58, 4.23], and 3.59 [95\% CI: 3.00, 4.18] kcal/mol respectively, with corresponding force RMSE values of 11.76 [95\% CI: 11.41, 12.12], 10.22 [95\% CI: 9.82, 10.68], 9.13 [95\% CI: 8.67, 9.70] kcal/mol$\cdot$ ${\text{Å}}^{-1}$; \texttt{espaloma-0.3} performed superiorly better for all three datasets.

\begin{figure}[tb]
    \centering
    \includegraphics[width=0.8\textwidth]{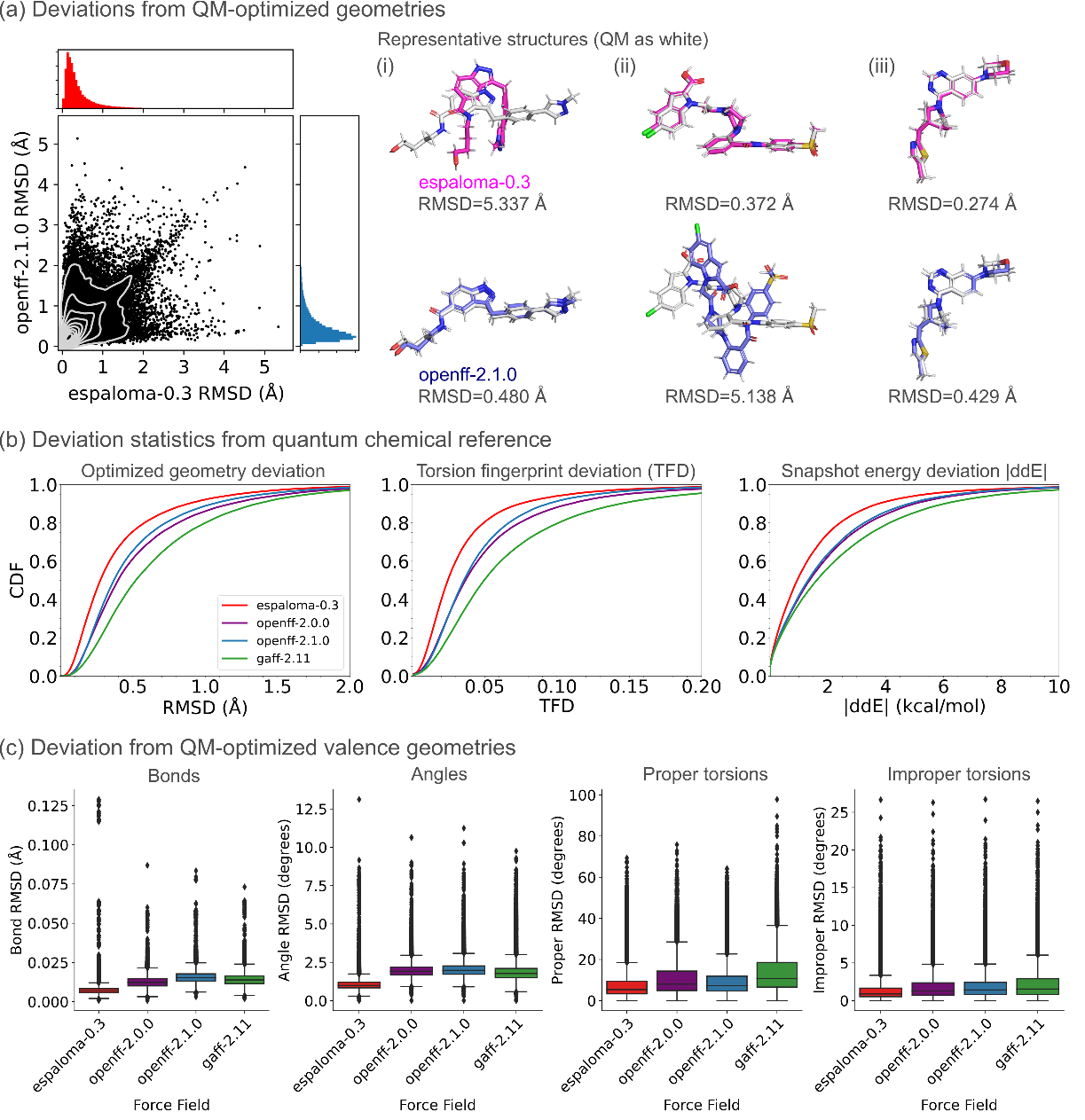}
    \caption{\label{fig:offopt}
    \textbf{\texttt{espaloma-0.3} preserves the location of quantum chemical energy minima.} 
    An industry standard benchmark of gas-phase QM-optimized geometries (the \texttt{OpenFF Industry Benchmark Season 1 v1.1}~\cite{damore2022collaborative} from QCArchive), comprising 9728 unique molecules and 73\,301 conformers, was used to compare the structures and energetics of conformers optimized using \texttt{espaloma-0.3}, \texttt{openff-2.0.0}~\cite{openff-2.0.0}, \texttt{openff-2.1.0}~\cite{openff-2.1.0}, and  \texttt{gaff-2.11}~\cite{gaff2} with respect to their QM-optimized geometries at the B3LYP-D3BJ/DZVP level of theory.
    (a) Representative scatter plot of root-mean-square deviation (RMSD) of atomic positions between \texttt{espaloma-0.3} and \texttt{openff-2.1.0}. The superposed structures between the QM-optimized (white) and MM-optimized geometries with the maximum RMSD obtained by (i) \texttt{espaloma-0.3}, (ii) \texttt{openff-2.1.0}, and (iii) the median RMSD of \texttt{espaloma-0.3} are shown.
    (b) The cumulative distribution functions of root-mean-square deviation (RMSD) of atomic positions, torsion fingerprint deviation (TFD) score, and relative energy differences (ddE) as described in a previous work~\cite{lim2020f1000} are reported.
    (c) Distributions of bond, angle, proper torsion, and improper torsion RMSD within each conformer with respect to its QM-optimized geometries are shown as quartile box plots. Lower values for all metrics indicate that the MM-optimized geometry is close to the quantum chemical reference structure.
    }
\end{figure}

\section{Espaloma force field preserves quantum chemical energy minima}
\label{sec:geoopt}

We next examined whether the ability of \texttt{espaloma-0.3} to quantitatively reproduce the quantum chemical equilibrium conformational energetics extends to an ability to qualitatively preserve the conformations of quantum chemical local energy minima---important for accurately representing geometries for phenomena like ligand binding docking studies, simulations, or free energy calculations.
To assess this, we used a standardized industry benchmark of gas-phase QM-optimized geometries (the \texttt{OpenFF Industry Benchmark Season 1 v1.1}\protect\footnote{\url{https://github.com/openforcefield/qca-dataset-submission/tree/master/submissions/2021-06-04-OpenFF-Industry-Benchmark-Season-1-v1.1}}~\cite{damore2022collaborative} obtained from QCArchive) to compare the structures and energetics of conformers optimized using \texttt{espaloma-0.3} and baseline force fields (\texttt{openff-2.0.0}, \texttt{openff-2.1.0}, and \texttt{gaff-2.11}) with respect to their QM-optimized geometries at the B3LYP-D3BJ/DZVP level of theory.
The dataset is a collection of drug-like molecules selected by industry partners of the Open Force Field Consortium and is representative of their current interests in chemical spaces, serving as an out-of-distribution test dataset.
It contains 9728 unique molecules and 73\,301 conformers after filtering out any quantum chemical calculation failures due to convergence issues and connectivity changes during geometry optimization.

As shown in {\bf Figure~\ref{fig:offopt} (a,b)}, the geometries and relative conformer energies with respect to their quantum chemical reference values showed better agreement with \texttt{espaloma-0.3} than with the baseline force fields---\texttt{openff-2.0.0}, \texttt{openff-2.1.0}, and \texttt{gaff-2.11}.
Additionally, the bonds, angles, and torsions in MM-optimized geometries obtained using \texttt{espaloma-0.3} show close agreement with quantum chemical values ({\bf Figure~\ref{fig:offopt} (c)}), resulting in an overall performance compatible or slightly better than the baseline force fields.
The bond outliers (>0.1~${\text{Å}}$) with \texttt{espaloma-0.3} arise from three sulfonamides connected to aliphatic carbons, comprising a total of 30 conformers---0.04\% of the conformers in the entire benchmark dataset---exhibiting $\sim$0.4~${\text{Å}}$ elongated S-N bond distances in the sulfonamide groups compared to the QM-optimized geometries ({\bf SI Figure~\ref{fig:offopt_outlier_si} (a)}). 
12 other molecules containing sulfonamide groups, excluding the bond RMSD outliers were found within the benchmark dataset with each molecular conformer featuring reasonable bond distances within the sulfonamide group ({\bf SI Figure~\ref{fig:offopt_outlier_si} (b)}).
However, the nitrogen geometry of pyrazoles and imidazoles substituted with sulfonamides became trigonal pyramidal when minimized with \texttt{espaloma-0.3}, rather than preserving a flat ring geometry and losing their sp2 hybridized features, as observed with QM-optimized geometries ({\bf SI Figure~\ref{fig:offopt_outlier_si} (c)}).
The angle outlier is also related to a sulfonamide but was a singleton of a non-druglike molecule containing a single conformer, with $\sim$40 degree deviation from its original QM-optimized geometry ({\bf SI Figure~\ref{fig:offopt_outlier_si} (a)}). 

Nonetheless, the degree of improvement of \texttt{espaloma-0.3} relative to \texttt{openff-2.0.0} is surprising and intriguing, considering that the Lennard-Jones parameters are transferred from \texttt{openff-2.0.0} and the overlap in the underlying \texttt{Optimization} and \texttt{TorsionDrive} datasets used for optimizing both force fields. This is notable, despite \texttt{espaloma-0.3} was trained on quantum chemical dataset comprising larger and broader chemical species.
\begin{figure}[tb]
    \centering
    
    \begin{subfigure}{0.198\textwidth}
        \centering
        \caption{Chi-square values}
        \includegraphics[width=1.0\linewidth]{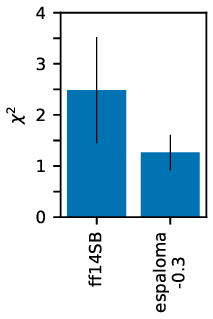}
    \end{subfigure}
    \hfill
    \begin{subfigure}{0.792\textwidth}
        \centering
        \caption{Comparison of estimated and experimental scalar couplings}
        \includegraphics[width=1.0\linewidth]{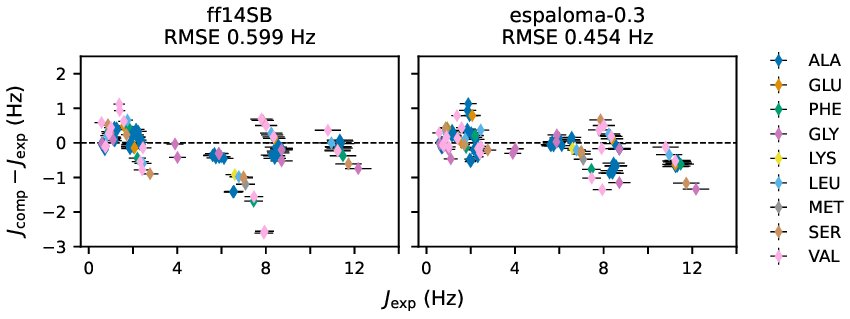}
    \end{subfigure}
    
    \caption{\label{fig:biopolymer}
    \textbf{\texttt{espaloma-0.3} reproduces experimental NMR scalar couplings of unstructured peptides better than well-established biomolecular force fields.}
    (a) $\chi^2$ values (lower is better) quantifying deviations of simulated NMR scalar couplings computed from 500 ns trajectories from experimental NMR measurements~\cite{graf2007structure, hagaram2010intrinsic}. 
    Error bars represent a 95\% confidence interval constructed from the critical values of a Student’s t distribution and the standard error of the mean across the NMR observables.
    (b) Comparison of the error in computed estimates of NMR scalar couplings versus experiment. Colors represent the identity of the amino acid associated with each scalar coupling. Horizontal error bars represent the estimate of the systematic error in the experimental scalar coupling, and vertical error bars represent the uncertainty due to the computed estimate (standard error of the mean across 3 replicates) and the uncertainty due to the experimental value (systematic error) added in quadrature.}
\end{figure}

\section{Espaloma force field reproduces experimental NMR observables for peptides}
\label{sec:biopolymer}

To quantitatively assess the ability of \texttt{espaloma-0.3} to model the intrinsic backbone preferences of amino acids, we performed MD simulations of thirteen short, unstructured peptides for which NMR observables have been experimentally measured~\cite{graf2007structure, hagaram2010intrinsic}.
The peptides are composed of 3 to 5 residues, uncapped, and have protonated C Termini due to the low pH of the NMR experiments.
Measured vicinal scalar couplings inform on the backbone dihedral preferences of these peptides.
Scalar couplings were computed from 500 ns trajectories using a Karplus model~\cite{karplus1963vicinal, hu1997determination, hennig2000determination, wirmer2002angular, ding2004protein, vogeli2007limits}, and agreement with experimental observables was quantified using a $\chi^2$ value.

Overall, \texttt{espaloma-0.3} produces closer agreement with experiment than \texttt{ff14SB}, as evidenced by the low $\chi^2$ value ({\textbf{Figure~\ref{fig:biopolymer} (a)}}).
With note, \texttt{ff14SB} tends exhibits closer agreement with experiments on amino acids with short side chains such as glycine and alanine ({\textbf{Figure~\ref{fig:biopolymer} (b)}}).
This is unsurprising as the backbone torsion parameters for \texttt{ff14SB} were tuned to reproduce the NMR scalar couplings for the alanine 5-mer peptide included in this dataset~\cite{ff14SB}.
However, \texttt{espaloma-0.3} tends to have closer agreement with experiments on more challenging amino acids with charged (e.g. lysine), bulky (e.g. methionine), or $\beta$-branched (e.g. valine) side chains, reflecting the transferability of Espaloma’s neural network parameters---which were trained on gas phase quantum chemistry data—to the condensed phase.
\begin{figure}[tb]
    \centering
    \includegraphics[width=1.0\textwidth]{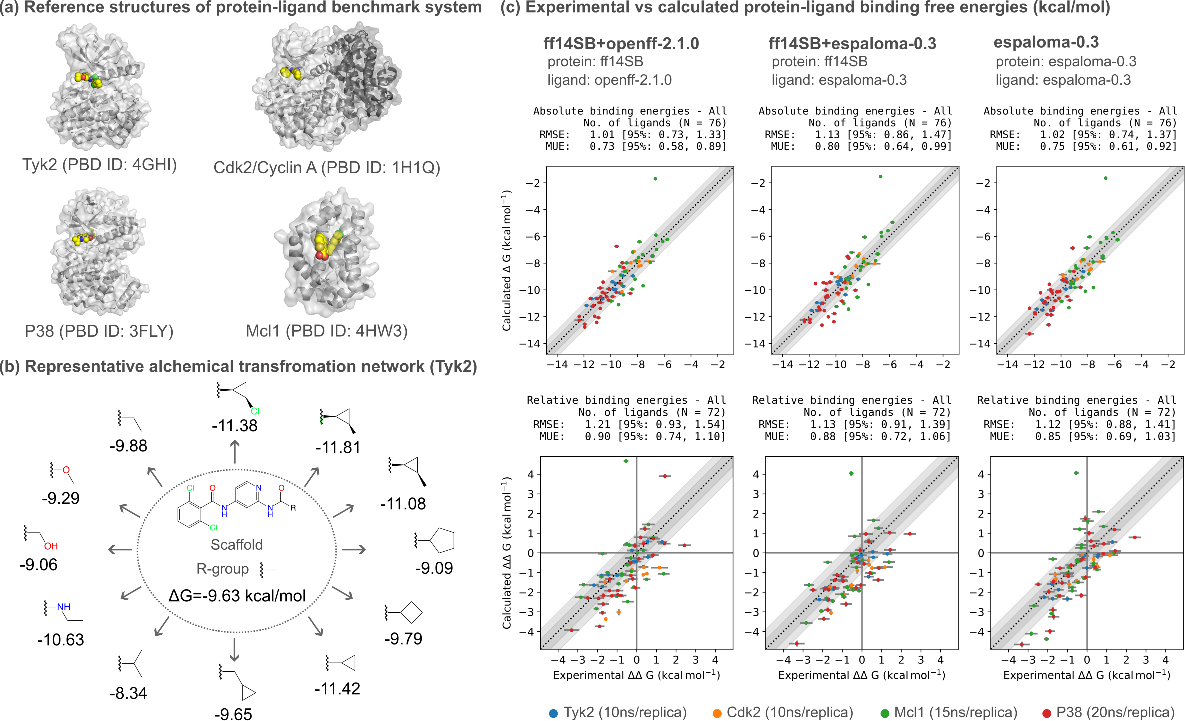}
    \caption{\label{fig:all}
    \textbf{\texttt{espaloma-0.3} can be used for accurate protein-ligand alchemical free energy calculations.}
    (a) Protein-ligand (PL) alchemical free energy calculations were calculated for Tyk2 (10 ns/replica), Cdk2 (10 ns/replica), Mcl1 (15 ns/replica), P38 (20 ns/replica) using a curated PL-benchmark dataset (see {\bf SI Section~\ref{sec:si_plbenchmark}}) which comprises 76 ligands in total. The PL structures used to setup the alchemical free energy calculations for each target system is shown. Here, we used Perses 0.10.1 relative free energy calculation infrastructure \cite{perses-0.10.1}, based on OpenMM 8.0.0 \cite{eastman2017openmm}, to assess the accuracy of \texttt{espaloma-0.3} and \texttt{openff-2.1.0}~\cite{openff-2.1.0} combined with Amber ff14SB force field \cite{ff14SB} for comparison. 
    (b) Schematic illustration of the alchemical ligand transformation network for Tyk2. The methyl R-group in the center is alchemically transformed into various R-groups. The binding free energy for each R-group is annotated alongside the respective R-groups.
    (c) The \texttt{openff-2.1.0}~\cite{openff-2.1.0} with protein parametrized with Amber ff14SB force field (\textbf{ff14SB+openff-2.1.0}) achieves an absolute free energy ($\Delta G$) RMSE of 1.01 [95\% CI: 0.73, 1.33] kcal/mol. The \texttt{espaloma-0.3} for predicting valence parameters and partial charges of small molecules combined with Amber ff14SB force field for proteins (\textbf{ff14SB+espaloma-0.3}) achieves an absolute free energy ($\Delta G$) RMSE of 1.13 [95\% CI: 0.86, 1.47] kcal/mol. Parametrizing small molecule and protein self-consistently with \texttt{espaloma-0.3} (\textbf{espaloma-0.3}) achieves absolute free energy ($\Delta G$) RMSE of 1.02 [95\% CI: 0.74, 1.37] kcal/mol which is comparable to those obtained by (\textbf{ff14SB+openff-2.1.0}) and (\textbf{ff14SB+espaloma-0.3}). All systems were solvated with TIP3P water~\cite{jorgensen1983comparison} and neutralized with 300 mM NaCl salt using Joung and Cheatham monovalent counterions~\cite{joung2008determination}. The light and dark gray regions depict the confidence bounds of 0.5 kcal/mol and 1.0 kcal/mol, respectively.}
\end{figure}
\begin{table}[tbp]
\centering
\resizebox{\textwidth}{!}{%
\begin{tabular}{cccccccccccc}
\toprule
   &
   &
   &
   &
   &
   &
  \multicolumn{6}{c}{\textbf{protien: ff14SB / ligand: openff-2.1.0}} \\
    &
    &
    &
    &
   &
   &
  \multicolumn{2}{c}{Relative ($\Delta\Delta G$)} &
  \multicolumn{4}{c}{Absolute ($\Delta G$)} \\ \cmidrule(r){7-8} \cmidrule(r){9-12}
  System &
  PDB ID &
  Compds &
  Edges &
  Range (kcal/mol) &
  ns/replica &
  RMSE &
  MUE &
  RMSE &
  MUE &
  R$^2$ &
  \small Spearman $\rho$ \\
  \midrule
  Tyk2 &
  4GIH &
  13 &
  12 &
  3.47 &
  10 &
  $0.54_{0.36}^{0.71}$ &
  $0.45_{0.28}^{0.62}$ &
  $0.50_{0.36}^{0.64}$ &
  $0.42_{0.27}^{0.57}$ &
  $0.80_{0.53}^{0.93}$ &
  $0.89_{0.75}^{0.96}$ \\
  Cdk2 &
  1H1Q &
  10 &
  9 &
  2.78 &
  10 &
  $1.43_{1.04}^{1.75}$ &
  $1.29_{0.80}^{1.67}$ &
  $0.74_{0.50}^{0.93}$ &
  $0.63_{0.41}^{0.86}$ &
  $0.48_{0.13}^{0.85}$ &
  $0.69_{0.30}^{0.92}$ \\
  Mcl1 &
  4HW3 &
  25 &
  24 &
  4.19 &
  15 &
  $1.50_{0.83}^{2.12}$ &
  $1.02_{0.63}^{1.55}$ &
  $1.36_{0.77}^{2.01}$ &
  $0.97_{0.66}^{1.41}$ &
  $0.50_{0.35}^{0.73}$ &
  $0.71_{0.57}^{0.86}$ \\
  P38 &
  3FLY &
  28 &
  27 &
  3.81 &
  20 &
  $1.06_{0.81}^{1.30}$ &
  $0.87_{0.65}^{1.09}$ &
  $0.90_{0.60}^{1.19}$ &
  $0.69_{0.50}^{0.92}$ &
  $0.57_{0.38}^{0.78}$ &
  $0.76_{0.63}^{0.89}$ \\
    &
    &
    &
    &
    &
   &
    &
    &
    &
    &
    &
    \\
  \hline

   &
   &
   &
   &
   &
   &
  \multicolumn{6}{c}{\textbf{protein: ff14SB / ligand: espaloma-0.3}} \\
    &
    &
    &
    &
   &
   &
  \multicolumn{2}{c}{Relative ($\Delta\Delta G$)} &
  \multicolumn{4}{c}{Absolute ($\Delta G$)} \\ \cmidrule(r){7-8} \cmidrule(r){9-12}
  System &
  PDB ID &
  Compds &
  Edges &
  Range (kcal/mol) &
  ns/replica &
  RMSE &
  MUE &
  RMSE &
  MUE &
  R$^2$ &
  \small Spearman $\rho$ \\
  \midrule
  Tyk2 &
  4GIH &
  13 &
  12 &
  3.47 &
  10 &
  $0.70_{0.34}^{0.98}$ &
  $0.52_{0.28}^{0.80}$ &
  $0.48_{0.29}^{0.65}$ &
  $0.37_{0.23}^{0.55}$ &
  $0.79_{0.49}^{0.95}$ &
  $0.89_{0.71}^{0.97}$ \\
  Cdk2 &
  1H1Q &
  10 &
  9 &
  2.78 &
  10 &
  $1.15_{0.85}^{1.44}$ &
  $1.05_{0.73}^{1.36}$ &
  $0.56_{0.32}^{0.74}$ &
  $0.46_{0.27}^{0.66}$ &
  $0.63_{0.27}^{0.92}$ &
  $0.80_{0.53}^{0.96}$ \\
  Mcl1 &
  4HW3 &
  25 &
  24 &
  4.19 &
  15 &
  $1.38_{0.90}^{1.96}$ &
  $1.06_{0.76}^{1.44}$ &
  $1.51_{0.90}^{2.15}$ &
  $1.08_{0.74}^{1.56}$ &
  $0.60_{0.42}^{0.80}$ &
  $0.77_{0.63}^{0.90}$ \\
  P38 &
  3FLY &
  28 &
  27 &
  3.81 &
  20 &
  $1.03_{0.81}^{1.26}$ &
  $0.82_{0.59}^{1.05}$ &
  $1.10_{0.86}^{1.32}$ &
  $0.88_{0.63}^{1.13}$ &
  $0.38_{0.11}^{0.64}$ &
  $0.62_{0.34}^{0.80}$ \\
    &
    &
    &
    &
    &
   &
    &
    &
    &
    &
    &
    \\
  \hline

   &
   &
   &
   &
   &
   &
  \multicolumn{6}{c}{\textbf{protein: espaloma-0.3 / ligand: espaloma-0.3}} \\
    &
    &
    &
    &
   &
   &
  \multicolumn{2}{c}{Relative ($\Delta\Delta G$)} &
  \multicolumn{4}{c}{Absolute ($\Delta G$)} \\ \cmidrule(r){7-8} \cmidrule(r){9-12}
  System &
  PDB ID &
  Compds &
  Edges &
  Range (kcal/mol) &
  ns/replica &
  RMSE &
  MUE &
  RMSE &
  MUE &
  R$^2$ &
  \small Spearman $\rho$ \\
  \midrule
  Tyk2 &
  4GIH &
  13 &
  12 &
  3.47 &
  10 &
  $0.67_{0.45}^{0.87}$ &
  $0.56_{0.35}^{0.76}$ &
  $0.46_{0.33}^{0.58}$ &
  $0.40_{0.28}^{0.53}$ &
  $0.81_{0.64}^{0.94}$ &
  $0.90_{0.79}^{0.97}$ \\
  Cdk2 &
  1H1Q &
  10 &
  9 &
  2.78 &
  10 &
  $0.84_{0.58}^{1.05}$ &
  $0.75_{0.51}^{0.99}$ &
  $0.63_{0.48}^{0.76}$ &
  $0.58_{0.41}^{0.74}$ &
  $0.47_{0.14}^{0.82}$ &
  $0.68_{0.41}^{0.90}$ \\
  Mcl1 &
  4HW3 &
  25 &
  24 &
  4.19 &
  15 &
  $1.44_{0.96}^{1.99}$ &
  $1.10_{0.76}^{1.50}$ &
  $1.40_{0.78}^{2.09}$ &
  $1.00_{0.67}^{1.43}$ &
  $0.56_{0.40}^{0.78}$ &
  $0.75_{0.63}^{0.88}$ \\
  P38 &
  3FLY &
  28 &
  27 &
  3.81 &
  20 &
  $1.02_{0.77}^{1.24}$ &
  $0.79_{0.56}^{1.04}$ &
  $0.91_{0.68}^{1.13}$ &
  $0.75_{0.57}^{0.95}$ &
  $0.47_{0.24}^{0.68}$ &
  $0.68_{0.49}^{0.82}$ \\
\bottomrule
\end{tabular}%
}

\caption{\textbf{Protein-ligand alchemical free energy calculation benchmarks show \texttt{espaloma-0.3} achieves high accuracy that is competitive to well-established force fields.} 
Here, we report several different metrics to assess the performance of the protein-ligand binding benchmark results including root mean square error (RMSE), mean unsigned error (MUE), the square of the correlation coefficient (R$^2$), and the Spearman’s rank correlation coefficient ($\rho$) along with 95\% CI for each metric. The initial PDB ID, number of compounds, number of edges (ligand transformations), the binding affinity range, and the simulation time per replica are reported in the table.
}
\label{tab:summary}
\end{table}

\section{Espaloma force field accurately describes protein-ligand binding free energies}
\label{sec:binding}

To evaluate \texttt{espaloma-0.3} for real-world drug discovery applications, we performed relative alchemical free energy calculations on a curated protein-ligand binding benchmark dataset, which was adopted from the Open Force Field protein-ligand benchmark dataset~\footnote{\url{https://github.com/openforcefield/protein-ligand-benchmark/tree/d3387602bbeb0167abf00dfb81753d8936775dd2}}(see {\bf SI Section~\ref{sec:si_plbenchmark}}).
We selected target systems from available datasets based on several criteria: firstly, we prioritized systems with ligands that can be effectively modeled to alleviate the potential sampling issues arising from poor initial ligand poses; secondly, we excluded systems with cofactors and ions near the ligand binding site to simplify the evaluation; thirdly, we considered systems with diverse structure-activity relationships, including ligand net charges, multiple R-group enumeration, and scaffold hopping.
As a result, we selected four well-studied protein-ligand binding benchmark systems.
The protein structures, ligand poses, and ligand transformation networks were manually curated to ensure the free energy benchmark was an accurate and reproducible assessment of force field accuracy.
\begin{itemize}
    \item \texttt{Tyk2} (PDB: \texttt{4GIH})~\cite{liang2013lead}, a non-receptor tyrosine-protine kinase, has therapeutic significance in inflammatory bowel diseases (IBD). 
    This particularly popular system has good convergence and served as a control experiment.
    \item \texttt{Cdk2} (PDB: \texttt{1H1Q})~\cite{davies2002structure}, a cyclin-dependent kinase, is involved in molecular pathology of cancer and is, therefore, a popular target for structure-based drug design.
    We use this system, complex with cyclin A, to test the capability to parametrize multiple protein subunits.
    \item \texttt{P38} (PDB: \texttt{3FLY})~\cite{labute2021optimizing} is a mitogen-activated protein (MAP) kinase which is a central component in signaling networks in mammalian cell types.
    This target is another well-studied system, but is expected to be more challenging compared to Tyk2 and Cdk2 because of the larger ligand transformations and exploration of structure-activity relationships with multiple R-groups from different scaffold positions.
    \item \texttt{Mcl1} (PDB: \texttt{4HW3})~\cite{friberg2013discovery} (myeloid cell leukemia 1) is a member of the Bcl-2 family of proteins, which is overexpressed in various cancers and promotes aberrant survival of tumor cells.
    This target entails all ligands with a net charge of -1 and includes scaffold hopping; thus, chosen to test the capability to simulate free energy calculations for charged ligands and scaffold hopping. 
\end{itemize}
Within each system, we benchmarked three approaches of parametrization to evaluate the accuracy of \texttt{espaloma-0.3} in modeling either the ligand alone or the entire protein-ligand complex:
\begin{itemize}
    \item \textbf{Protein: ff14SB / Ligand: openff-2.1.0 (ff14SB+openff-2.1.0)}: 
    As a baseline, we parametrize the ligand region using a well-established small molecule force field \texttt{openff-2.1.0}~\cite{openff-2.1.0} and use the Amber ff14SB~\cite{ff14SB} to parametrize the protein.
    \item \textbf{Protein: ff14SB / Ligand: espaloma-0.3 (ff14SB+espaloma-0.3)}:
    We parametrize the ligand region using \texttt{espaloma-0.3} and use the Amber ff14SB~\cite{ff14SB} to parametrize the protein. We only parametrize the ligand region with \texttt{espaloma-0.3} to provide a head-to-head comparison with \texttt{openff-2.1.0}.
    \item \textbf{Protein: espaloma-0.3 / Ligand: espaloma-0.3 (espaloma-0.3)}:
    We apply \texttt{espaloma-0.3} to both the \textit{ligand} and  \textit{protein} regions of the system.
    This is to test the capability of \texttt{espaloma-0.3} to entirely replace the force field parametrization pipeline. 
    Instead of using two separate force fields for small molecules and proteins, each developed independently, we aim to apply a self-consistently developed force field that covers different chemical domains.
\end{itemize}
As our training dataset does not yet include water and ions, all systems were solvated with TIP3P water \cite{jorgensen1983comparison} and neutralized with the Joung and Cheatham monovalent counterions \cite{joung2008determination}. 
The perses 0.10.1 infrastructure~\cite{perses-0.10.1} was used to perform the alchemical protein-ligand binding free energy calculations (see {\bf SI Section~\ref{sec:si_perses}}).

In {\bf Figure~\ref{fig:all}} and {\bf Table~\ref{tab:summary}}, we illustrate that \texttt{espaloma-0.3}, where both the protein and ligand are parametrized self-consistently, has \textit{comparable} protein-ligand binding free energy performance with \texttt{ff14SB+openff-2.1.0}.
\texttt{espaloma-0.3} achieves absolute ($\Delta G$) and relative ($\Delta\Delta G$) free energy RMSE of 1.02 [95\% CI: 0.74, 1.37]  kcal/mol and 1.12 [95\% CI: 0.88, 1.41] kcal/mol, respectively. Correspondingly, the $\Delta G$ and $\Delta\Delta G$ RMSE for \texttt{ff14SB+openff-2.1.0} were 1.01 [95\% CI: 0.73, 1.33] kcal/mol and 1.21 [95\% CI: 0.93, 1.54] kcal/mol, respectively.
Although, the reported error and correlation statistics have overlapping confidence intervals, these results are encouraging as \texttt{espaloma-0.3} demonstrates its capability to cover different chemical domains, which traditional force fields have struggled for decades and have not accomplished.

Notably, a large outlier for the Mcl1 system for all three cases was observed as shown in {\bf Figure~\ref{fig:all}}. The problematic ligand transformation and the initial ligand pose is illustrated in {\bf SI Figure~\ref{fig:mcl1_outlier}}. The relative binding affinity $\Delta\Delta G$ computed with \texttt{ff14SB+espaloma-0.3} was 4.05 kcal/mol ({\bf Figure~\ref{fig:all} (b)}). However, we found that the error can be reduced to 2.60 kcal/mol when the alchemical binding free energy calculation was performed from a flipped binding pose, which is in better agreement with the experimental difference ({$\Delta\Delta G$} = -0.54 kcal/mol).

We also conducted another set of free energy calculations for the four target systems, each with three parametrization approaches ({\bf SI Figure~\ref{fig:converge}}).
In most cases, the absolute ($\Delta G$) and relative ($\Delta\Delta G$) binding free energies from the two independent trials were within 1.0 kcal/mol, demonstrating reasonable reproducibility; except for P38, which tends to be a more challenging target for the free energy calculations to reproduce.

It is worth noting that the ligands from the protein-ligand binding benchmark dataset are highly dissimilar to the molecules used in developing \texttt{espaloma-0.3}, with a maximum Tanimoto similarity of 0.5 between the two sources, suggesting the high generalizability of Espaloma ({\bf SI Figure~\ref{fig:tanimoto}}).

\paragraph{Regularization and larger training dataset significantly improve performance}

\label{sec:binding_firstgen}

To assess the impact of dataset scale and the regularization procedures introduced here for training \texttt{espaloma-0.3}, we compared the protein-ligand binding free energy calculations using the first-generation Espaloma force field (\texttt{0.2.2})~\cite{D2SC02739A}, which was trained on a limited quantum chemical dataset and without regularization compared to \texttt{0.3}.
The free energy calculations were conducted for all four target systems and were prepared similarly to those described above. 
In {\bf SI Figure ~\ref{fig:cdk2}}, \texttt{espaloma-0.2.2} significantly underperforms compared to \texttt{espaloma-0.3} for the Cdk2 system due to a large outlier. 
\texttt{espaloma-0.2.2} also demonstrates lesser performance on the Tyk2 system, as illustrated in {\bf SI Figure~\ref{fig:tyk2}}. 
Importantly, the protein-ligand binding free energy calculations were unstable for Mcl1 and P38, with many of the ligand transformations being suspended during the simulation. These results indicate that \texttt{espaloma-0.3}, trained on an extensive quantum chemical dataset and with an improved training strategy, has resulted in the development of a robust and stable Espaloma force field.

\section{Espaloma force field produces stable long-time molecular dynamics}
\label{sec:vanilla}

Recent benchmarks of machine learned force fields demonstrated that many of these potentials are accurate but cannot produce stable molecular dynamics simulations~\cite{fu2022forces}.
To assess whether \texttt{espaloma-0.3} was sufficiently stable and robust for general use in molecular dynamics simulations, we performed multiple replicates of a microsecond MD simulation of a solvated protein-ligand complex (Tyk2 complexed with ligand \#1, {\bf SI Figure~\ref{fig:tyk2}}) and monitored the ligand and protein root-mean square deviation (RMSD) profiles, as shown in {\bf SI Figure~\ref{fig:tyk2_vanilla}}. 
The simulations parametrized with \texttt{espaloma-0.3} remained comparably stable to those generated with \texttt{ff14SB+openff-2.1.0}, with both protein and ligand RMSD generally remaining below 2.0 ${\text{Å}}$.

\section{Discussion}
\label{sec:conclusion}

In this study, we introduced an enhanced graph neural network approach to rapidly construct a new generation of accurate, robust, and generalizable machine-learned MM force field, \texttt{espaloma-0.3}, capable of fine-tuning and extending to new chemical domains of interest. 
The newly developed force field captures both quantitative and qualitative behavior of quantum chemical conformational energetics for a wide range of chemical species. 
As a result, it not only recapitulates quantum chemical conformational energetics and geometries, but it also reproduces experimental NMR observables for peptides, leading to accurate predictions of protein-ligand binding free energies when both the protein and ligand are self-consistently parametrized with \texttt{espaloma-0.3}. 
We hope this work will lay the foundations to inspire the design of new generations of machine learning-empowered molecular mechanics force fields that can self-consistently describe the wide chemical domains relevant to biomolecular modeling and drug discovery.

\paragraph{An open chemically and conformationally diverse quantum chemical dataset was curated to construct \texttt{espaloma-0.3}}

In this paper, we have curated a high-quality open dataset covering chemical spaces and conformational regions of interest to biomolecular modeling, including small molecules, peptides, and RNA. 
We demonstrated how our enhanced Espaloma framework can scale to foundational quantum chemical datasets, enabling the achievement of a stable machine-learned MM force field. 
We released this dataset along with our implementation in the hope that this will enable the community to further optimize MM force fields by building on this dataset, or fine-tuning the \texttt{espaloma-0.3} model with additional data much the way foundational large language models (LLMs) can be fine-tuned to perform better on domain tasks of interest.

\paragraph{\texttt{espaloma-0.3} quantitatively and qualitatively recapitulates quantum chemical conformational energy landscapes}

We demonstrated that current force fields typically exhibit considerable disagreement with quantum chemical calculations in terms of reproducing conformational energies and forces (Table~\ref{tab:rmse}). 
With carefully crafted training and regularization strategies, we show that \texttt{espaloma-0.3} not only quantitatively agrees more closely with quantum chemical conformational energetics for a wide variety of chemical species, but also behaves qualitatively similarly with quantum chemistry, even in low data regimes ({\bf SI Figure~\ref{fig:train_size}}).
Although \texttt{espaloma-0.3} poses a challenge in preserving the quantum chemical energy minima for some sulfonamide groups ({\bf SI Figure~\ref{fig:offopt_outlier_si}}), more rigorous hyperparemeter tuning of the Espaloma framework may help resolve this problem, especially adjusting the weights for each loss component, as we find this to be sensitive to the overall performance.

\paragraph{Chemical diversity and high-energy conformers are important for accurately capturing quantum chemical energies and forces with Espaloma}

The cross-validation experiment ({\bf SI Figure~\ref{fig:leave_out}}), in which Espaloma is trained without certain categories of chemical species (small molecules, peptides, or RNA), suggests that quantum chemical datasets with broad chemical coverage---specifically, the SPICE-Pubchem (small molecules) dataset---can perceive and extrapolate the chemical environments for out-of-distributed chemical domains. 
A lack of chemical diversity leads to large quantum chemical force errors, whereas reproducing energies is easier ({\bf SI Figure~\ref{fig:leave_out} (a)}). 
Similarly, cross-validating certain dataset classes (single-point energies generated by MD [\texttt{Dataset}], optimization trajectories of enumerated conformers [\texttt{OptimizationDataset}], or one-dimensional torsion drives [\texttt{TorsionDriveDataset}]) suggests that high-energy conformers may be important to accurately capture the quantum chemical energies and forces with Espaloma and other machine learning-based methods ({\bf SI Figure~\ref{fig:leave_out} (b)})~\cite{wang2023denoise}. 
The quantum chemical forces of peptide datasets, including local energy minima conformers (Pepconf-Opt dataset from [\texttt{OptimizationDataset}]), were poorly reproduced when trained without datasets storing relatively high energy conformers (SPICE-Dipeptide dataset from [\texttt{Dataset}]).

\paragraph{\texttt{espaloma-0.3} can be easily extended to other chemical spaces of interest}
The chemical space covered by an Espaloma force field can easily be extended to spaces highly relevant in other areas of biomolecular modeling, such as lipids, DNA, and glycans, by simply augmenting the quantum chemical dataset used in training. 
In constructing \texttt{espaloma-0.3}, we demonstrated that this approach easily scales to 1.1 million energies and forces, representing nearly 17\,000 chemical species, in less than a single GPU-day.
Because loss function is easily parallelizable, this approach should scale gracefully to much larger datasets by simply distributing gradient computation across multiple GPUs, enabling rapid parametrization on much larger datasets or extension to new chemical domains of interest.

\paragraph{Espaloma offers a modular and extensible approach to building MM force fields}
Since the Espaloma architecture and loss function are modular~\cite{D2SC02739A} and, as demonstrated here, new force fields can be trained in a single GPU-day, Espaloma offers the opportunity to rapidly explore different MM functional forms. 
For example, many molecular mechanics simulation packages support atom-pair specific 1--4 Lennard-Jones and electrostatic parameters, alternative Lennard-Jones mixing rules, or alternative functional forms for van der Waals treatment.
Of particular interest are Class~II force fields~\cite{dauber2019biomolecular,hagler2019force}, where higher-order couplings between valence terms are introduced to reproduce the bond and angle vibrations more accurately---while the combinatorial explosion of these terms presents a problem for atom type based force fields, Espaloma does not suffer from the same issue and may provide a robust way to parametrize these force fields.



\paragraph{Espaloma fit to condensed-phase properties can further improve accuracy}
While we have demonstrated the ability to create a force field capable of reproducing NMR observables for peptides and predicting accurate protein-ligand binding free energies solely from fitting to quantum chemical data, further assessment is needed to confirm its ability to accurately reproduce condensed-phase properties. 
Since non-bonded interactions are generally optimized to fit condensed-phase properties, training against these properties may be necessary. 
An earlier study has shown that optimizing against condensed-phase mixture properties, rather than properties of pure systems, is better suited to improve force field accuracy for biomolecular systems~\cite{boothroyd2022improving}. 
The end-to-end differentiable nature of Espaloma makes it possible to employ reweighting approaches to directly fit to experimental free energies or thermodynamics~\cite{wieder2021fitting,wieder2021teaching,setiadi2023tuning} or other thermophysical properties~\cite{boothroyd2022improving}.
This could either be done directly during fitting or during a second-stage fine-tuning procedure that adapts an Espaloma force field to specific applications of interest.

\paragraph{Quantifying force field uncertainty could help generate more robust force fields}
One of the challenges in force field development is quantifying the contribution of errors in the force field to predicted quantities.
While statistical error is generally reported, this systematic force field error is frequently the major source of error in biomolecular simulations.
In recent years, several approaches have emerged to quantify uncertainty in deep learning methods, including mean-variance estimation, Bayesian methods, and ensemble methods~\cite{coley2020jcim,gawlikowski2022survey, wang2021stochastic}. 
Employing these methods to propagate force field uncertainty into predicted free energies and physical properties could enable Espaloma to provide a quantitative assessment of force field uncertainty.
With a better understanding of how this uncertainty propagates to task predictions, we envision that uncertainty-based active learning~\cite{smith2018less} (with possibly machine learning surrogate~\cite{wang2023spatial}) or adversarial attacks~\cite{gomez2021nanturecomm} could be employed to identify the most valuable new data to be generated in future efforts to train more robust Espaloma force fields.

\section{Data availability}
\label{sec:data}
The raw quantum chemical datasets downloaded from QCArchive is deposited in Zenodo (\url{https://zenodo.org/record/8148817}).
The pre-processed input data used to train \texttt{espaloma-0.3} is deposited in Zenodo (\url{https://zenodo.org/record/8150601}). 
The QM- and MM-minimized structures used for the small molecule geometry benchmark study is deposited in Zenodo (\url{https://doi.org/10.5281/zenodo.8378216}).
\section{Code availability}
\label{sec:code}
The Python code to download the quantum chemical data from QCArchive is available from \url{https://github.com/choderalab/download-qca-datasets}.
The scripts used to train and evaluate \texttt{espaloma-0.3} is available from \url{https://github.com/choderalab/refit-espaloma}.
The scripts used to perform the small molecule geometry benchmark is available from \url{https://github.com/choderalab/geometry-benchmark-espaloma}.
The curated protein-ligand benchmark dataset can be found from \url{https://github.com/kntkb/protein-ligand-benchmark-custom}, and the scripts to perform and analyze the alchemical protein-ligand binding affinity calculation with Perses is available from \url{https://github.com/choderalab/pl-benchmark-espaloma-experiment}.
The scripts used to perform the MD simulation of Tyk2 protein-ligand system is available from \url{https://github.com/choderalab/vanilla-espaloma-experiment}.
These python codes are also summarized in \url{https://github.com/choderalab/espaloma-0.3.0-manuscript}.
The code used for the peptide benchmark study is available from \url{https://github.com/openforcefield/proteinbenchmark}.
\section{Author Contributions}
Conceptualization: KT, YW, JDC;
Methodology: KT, YW;
Investigation: KT, PKB, CEC;
Software: KT, YW, IP, MMH, HM, CRI;
Writing -- Original Draft: KT, YW;
Writing -- Review \& Editing: KT, IP, PKB, CEC, AJF, MMH, HM, CRI, AMN, AMP, MRS, DLM, JDC, YW;
Funding Acquisition: JDC;
Resources: JDC;
Supervision: JDC, YW.

\section{Disclosures}
J.D.C.\ is a current member of the Scientific Advisory Board of OpenEye Scientific Software, Redesign Science, Ventus Therapeutics, and Interline Therapeutics, and has equity interests in Redesign Science and Interline Therapeutics. 
The Chodera laboratory receives or has received funding from multiple sources, including the National Institutes of Health, the National Science Foundation, the Parker Institute for Cancer Immunotherapy, Relay Therapeutics, Entasis Therapeutics, Silicon Therapeutics, EMD Serono (Merck KGaA), AstraZeneca, Vir Biotechnology, Bayer, XtalPi, Interline Therapeutics, the Molecular Sciences Software Institute, the Starr Cancer Consortium, the Open Force Field Consortium, Cycle for Survival, a Louis V. Gerstner Young Investigator Award, and the Sloan Kettering Institute. A complete funding history for the Chodera lab can be found at \url{http://choderalab.org/funding}.
Y.W.\ has limited financial interests in Flagship Pioneering, Inc. and its subsidiaries.
M.R.S.\ is an Open Science Fellow with Psivant Sciences and consults for Relay Therapeutics.
D.L.M.\ serves on the scientific advisory boards of Anagenex and OpenEye Scientific Software, Cadence Molecular Sciences, and is an Open Science Fellow with Psivant.

\section{Acknowledgements}
The authors thank OpenEye Scientific Software for providing a free academic license to the OpenEye Toolkits. 
The authors are also grateful for the OpenFF R01 grant and to all those that provided feedback on versions of the manuscript, including (but not limited to) Michael K. Gilson, Demetri Moustakas, Yutong Zhao, Tristan Croll, Domenico Bonanni, Timothy Bernat, and Mark E. Tuckerman.
This research was carried out on high performance computing resources at Memorial Sloan Kettering Cancer Center and the Washington Square and Abu Dhabi campuses of New York University.
This work used Bridges-2 at Pittsburgh Supercomputing Center through allocation Accelerate ACCESS BIO230106 from the Advanced Cyberinfrastructure Coordination Ecosystem: Services \& Support (ACCESS) program.

\section{Funding}
Y.W.\ acknowledges support from the Schmidt Science Fellowship, in partnership with the Rhodes Trust, and the Simons Center for Computational Physical Chemistry at New York University.
J.D.C.\ acknowledges support from NIH grant P30 CA008748, NIH grant R01 GM132386, NIH grant R01 GM121505, and the Sloan Kettering Institute.
M.R.S.\ acknowledges support from National Science Foundation grants \#2138259, \#2138286, \#2138307, \#2137603, and \#2138296.
D.L.M.\ appreciates support from the NIH grant R35GM148236 and R01GM132386.
P.K.B.\ appreciates financial support from the NIH NIGMS R01GM132386.

\bibliography{reference.bib}
\newpage
\appendix
\title{Appendix: Machine-learned molecular mechanics force field for the simulation of protein-ligand systems and beyond.}
\maketitle

\section{Code dependencies}
\label{sec:si_code}
Core dependencies include a modified version of Espaloma 0.3.0 release~\cite{D2SC02739A} (\url{https://github.com/choderalab/espaloma/tree/4c6155b72d00ce0190b3cb551e7e59f0adc33a56}), PyTorch 1.1.2~\cite{NEURIPS2019_9015}, Deep Graph Library 0.9.0~\cite{wang2019deep}, and Open Force Field Toolkit 0.10.6~\cite{openfftoolkit-0.10.6}, to refit and evaluate the espaloma model. A modified version of Openmmforcefields 0.11.0~\cite{openmmforcefields} (\url{https://github.com/kntkb/openmmforcefields/tree/6d2c3dcd33d9800a32032d28b6b2dca92f348a43}) was used to run all the relative alchemical protein-ligand binding free energy calculations with Perses 0.10.1 infrastructure~\cite{perses-0.10.1}. Espaloma 0.2.4 release and a modified version of Espaloma 0.3.0 was used to parametrize small molecules with \texttt{espaloma-0.2.2} and \texttt{espaloma-0.3}, respectively. A modified version of Perses 0.10.1 (\url{https://github.com/kntkb/perses/tree/0d069fc1cf31b8cce1ae7a1482c3fa46bc1382d2}) was used to self-consistently parametrize both small molecules and proteins with \texttt{espaloma-0.3}.
 A modified version of cinnabar 0.3.0~\cite{arsenic} (\url{https://github.com/kntkb/cinnabar/tree/de7bc6623fb25d75848aa1c9f538b77cd02a4b01}) was used to support arbitrary tick frequency when plotting $\Delta G$ and $\Delta\Delta G$ plots.
\section{MolSII QCArchive quantum chemical datasets}
\label{sec:si_dataset}
The Python code used to download the quantum chemical (QC) datasets from the MolSSI QCArchive~\cite{qcsubmit} is available at \url{https://github.com/choderalab/download-qca-datasets}. The QC datasets utilized in this study were obtained from various workflows implemented in the QCArchive ecosystem, including \texttt{Dataset}, \texttt{OptimizationDataset}, and \texttt{TorsionDriveDataset} generated at the B3LYP-D3BJ/DZVP level of theory. This level of theory was chosen to maintain consistency with the Open Force Field Consortium~\cite{parsley,sage}, and it is expected to balance computational efficiency and accuracy in reproducing conformations generated by higher-level theories~\cite{pavan2022openffbench}.

The QC datasets in {\bf Table~\ref{tab:rmse}} are composed of the following datasets deposited in QCArchive and annotated based on their respective categories.

\paragraph{Small molecules}
\begin{itemize}
    \item \textbf{SPICE-Pubchem}~\cite{eastman2023spice} 
    \footnote{Source: \url{https://github.com/openforcefield/qca-dataset-submission/tree/master/submissions/2021-11-08-QMDataset-pubchem-set1-single-points}}
    \footnote{Source: \url{https://github.com/openforcefield/qca-dataset-submission/tree/master/submissions/2021-11-08-QMDataset-pubchem-set2-single-points}}
    \footnote{Source: \url{https://github.com/openforcefield/qca-dataset-submission/tree/master/submissions/2021-11-09-QMDataset-pubchem-set3-single-points}}
    \footnote{Source: \url{https://github.com/openforcefield/qca-dataset-submission/tree/master/submissions/2021-11-09-QMDataset-pubchem-set4-single-points}}
    \footnote{Source: \url{https://github.com/openforcefield/qca-dataset-submission/tree/master/submissions/2021-11-09-QMDataset-pubchem-set5-single-points}}
    \footnote{Source: \url{https://github.com/openforcefield/qca-dataset-submission/tree/master/submissions/2021-11-09-QMDataset-pubchem-set6-single-points}}
    is a collection of \texttt{Dataset} that contains a comprehensive and diverse collection of small, drug-like molecules obtained from Pubchem~\cite{kim2023pubchem}. It includes atoms within the range of 3 to 50, including hydrogens, and encompasses the elements of Br, C, Cl, F, H, I, N, O, P, and S.
    
    \item \textbf{SPICE-DES-Monomers}~\cite{eastman2023spice} 
    \footnote{Source: \url{https://github.com/openforcefield/qca-dataset-submission/tree/master/submissions/2021-11-15-QMDataset-DES-monomers-single-points}}
    is a \texttt{Dataset}, sourced from DES370K~\cite{DES370K}, consists of  small molecules (up to 22 atoms) chosen to cover a wide range of chemical space, including the elements of Br, C, Cl, F, H, I, N, O, P, and S.
    
    \item \textbf{Gen2-Opt} 
    \footnote{Source: \url{https://github.com/openforcefield/qca-dataset-submission/tree/master/submissions/2020-03-20-OpenFF-Gen-2-Optimization-Set-1-Roche}}
    \footnote{Source: \url{https://github.com/openforcefield/qca-dataset-submission/tree/master/submissions/2020-03-20-OpenFF-Gen-2-Optimization-Set-2-Coverage}}
    \footnote{Source: \url{https://github.com/openforcefield/qca-dataset-submission/tree/master/submissions/2020-03-20-OpenFF-Gen-2-Optimization-Set-3-Pfizer-Discrepancy}}
    \footnote{Source: \url{https://github.com/openforcefield/qca-dataset-submission/tree/master/submissions/2020-03-20-OpenFF-Gen-2-Optimization-Set-4-eMolecules-Discrepancy}}
    \footnote{Source: \url{https://github.com/openforcefield/qca-dataset-submission/tree/master/submissions/2020-03-20-OpenFF-Gen-2-Optimization-Set-5-Bayer}} 
    is a collection of \texttt{OptimizationDataset} that contains  drug-like molecules used for the parametrization of the OpenFF 1.2.0 ("Parsley")~\cite{parsley} small molecule force field developed by the Open Force Field Consortium. This dataset is one of the datasets used to generate the first generation espaloma force field, \texttt{espaloma-0.2.2}.

    \item \textbf{Gen2-Torsion} 
    \footnote{Source: \url{https://github.com/openforcefield/qca-dataset-submission/tree/master/submissions/2020-03-12-OpenFF-Gen-2-Torsion-Set-1-Roche}}
    \footnote{Source: \url{https://github.com/openforcefield/qca-dataset-submission/tree/master/submissions/2020-03-23-OpenFF-Gen-2-Torsion-Set-1-Roche-2}}
    \footnote{Source: \url{https://github.com/openforcefield/qca-dataset-submission/tree/master/submissions/2020-03-12-OpenFF-Gen-2-Torsion-Set-2-Coverage}}
    \footnote{Source: \url{https://github.com/openforcefield/qca-dataset-submission/tree/master/submissions/2020-03-23-OpenFF-Gen-2-Torsion-Set-2-Coverage-2}}
    \footnote{Source: \url{https://github.com/openforcefield/qca-dataset-submission/tree/master/submissions/2020-03-12-OpenFF-Gen-2-Torsion-Set-3-Pfizer-Discrepancy}}
    \footnote{Source: \url{https://github.com/openforcefield/qca-dataset-submission/tree/master/submissions/2020-03-23-OpenFF-Gen-2-Torsion-Set-3-Pfizer-Discrepancy-2}}
    \footnote{Source: \url{https://github.com/openforcefield/qca-dataset-submission/tree/master/submissions/2020-03-12-OpenFF-Gen-2-Torsion-Set-4-eMolecules-Discrepancy}}
    \footnote{Source: \url{https://github.com/openforcefield/qca-dataset-submission/tree/master/submissions/2020-03-23-OpenFF-Gen-2-Torsion-Set-4-eMolecules-Discrepancy-2}}
    \footnote{Source: \url{https://github.com/openforcefield/qca-dataset-submission/tree/master/submissions/2020-03-12-OpenFF-Gen-2-Torsion-Set-5-Bayer}}
    \footnote{Source: \url{https://github.com/openforcefield/qca-dataset-submission/tree/master/submissions/2020-03-26-OpenFF-Gen-2-Torsion-Set-5-Bayer-2}}
    \footnote{Source: \url{https://github.com/openforcefield/qca-dataset-submission/tree/master/submissions/2020-03-12-OpenFF-Gen-2-Torsion-Set-6-supplemental}}
    \footnote{Source: \url{https://github.com/openforcefield/qca-dataset-submission/tree/master/submissions/2020-03-26-OpenFF-Gen-2-Torsion-Set-6-supplemental-2}} 
    is a collection \texttt{TorsionDriveDataset} that contains torsion scans of drug-like molecules which is part of the dataset used for the parametrization of the OpenFF 2.0.0 ("Sage")~\cite{sage} small molecule force field developed by the Open Force Field Consortium. 
\end{itemize}

\paragraph{Peptides}
\begin{itemize}
    \item \textbf{SPICE-Dipeptide}~\cite{eastman2023spice} 
    \footnote{Source: \url{https://github.com/openforcefield/qca-dataset-submission/tree/master/submissions/2021-11-08-QMDataset-Dipeptide-single-points}} 
    is a \texttt{Dataset} that contains a broad coverage of the possible dipeptides capped with ACE and NME groups formed by the 20 natural amino acids and their common protonation variants. This includes two forms of CYS (neutral or negatively charged), two forms of GLU (neutral or negatively charged), two forms of ASP (neutral or negatively charged), two forms of LYS (neutral or positively charged), and three forms of HIS (neutral forms with a hydrogen on either ND1 or NE2, and a positively charged form with hydrogens on both). 
    
    \item \textbf{Pepconf-Opt} 
    \footnote{Source: \url{https://github.com/openforcefield/qca-dataset-submission/tree/master/submissions/2020-10-26-PEPCONF-Optimization}} 
    is a \texttt{OptimizationDataset} that contains short peptides, including capped, cyclic, and disulfide-bonded peptides originally sourced from Prasad et al.~\cite{prasad2019pepconf} and regenerated by the Open Force Field Consortium. In this study, the \texttt{default-dlc} QC specification was utilized, differing from the one used in the first generation espaloma force field (\texttt{espaloma-0.2.2})~\cite{D2SC02739A}, leading to improved chemical convergence. 

    \item \textbf{Protein-torsion} 
    \footnote{Source: \url{https://github.com/openforcefield/qca-dataset-submission/tree/master/submissions/2021-11-18-OpenFF-Protein-Dipeptide-2D-TorsionDrive}} 
    \footnote{Source: \url{https://github.com/openforcefield/qca-dataset-submission/tree/master/submissions/2022-02-10-OpenFF-Protein-Capped-1-mer-Sidechains}} 
    \footnote{Source: \url{https://github.com/openforcefield/qca-dataset-submission/tree/master/submissions/2022-05-30-OpenFF-Protein-Capped-3-mer-Backbones}} 
    \footnote{Source: \url{https://github.com/openforcefield/qca-dataset-submission/tree/master/submissions/2023-02-06-OpenFF-Protein-Capped-3-mer-Omega}} 
    is a collection of \texttt{TorsionDriveDataset} that contains various torsion scans of polypeptides (capped 1-mers and capped 3-mers)  generated by the Open Force Field Consortium for the OpenFF 3.x ("Rosemary") force field~\cite{rosemary}. These torsion scans cover $\chi_1$ and $\chi_2$ angles in the rotatable side chains, as well as $\phi$, $\psi$, and $\omega$ angles in the backbones.
\end{itemize}

\paragraph{RNA}
\begin{itemize}
    \item \textbf{RNA-Diverse} 
    \footnote{Source: \url{https://github.com/openforcefield/qca-dataset-submission/tree/master/submissions/2022-07-07-RNA-basepair-triplebase-single-points}} 
    is a \texttt{Dataset} that contains comprehensive and diverse collection of experimental RNA structures. It includes 138 base pair structures and 295 base triple structures sourced from the Nucleic Acid Database~\cite{nucleic-acid-database}. Additionally, the dataset contains 4056 representative trinucleotide structures obtained from the RNA Structure Atlas website~\cite{rna-structure-atlas}, where the experimentally observed internal and hairpin loop motifs, as well as junction loops of representative sets of RNA 3D Structures with an X-ray resolution cutoff of 2.5~\AA, were segmented into all possible trinucleotide permutations, resulting in 64 unique molecules. These trinucleotide structures are capped with O5' hydroxyl groups at the 5' end and clustered to select the representative structures. For the espaloma refitting experiment, only the trinucleotides were utilized.

    \item \textbf{RNA-Trinucleotide} 
    \footnote{Source: \url{https://github.com/openforcefield/qca-dataset-submission/tree/master/submissions/2022-10-21-RNA-trinucleotide-single-points}} 
    is a \texttt{Dataset} that provides a broader and more diverse structural coverage of trinucleotides compared to the \textbf{RNA-Diverse} dataset.

    \item \textbf{RNA-Nucleoside} 
    \footnote{Source: \url{https://github.com/openforcefield/qca-dataset-submission/tree/master/submissions/2023-03-09-RNA-nucleoside-single-points}} 
    is a \texttt{Dataset} that comprises a comprehensive and diverse collection of nucleosides (adenosine, guanosine, cytidine, and uridine) without O5' hydroxyl atoms. These nucleosides are generated using 500 K implicit solvent MD and torsion scanning on N-glycosidic bond ($\chi$ torsion) that connects the base and sugar, resulting in diverse sugar pucker conformations and extensive coverage of $\chi$ torsions.
\end{itemize}

\section{Small molecule geometry optimization}
\label{sec:si_geoopt}
The Python code used to benchmark the small molecule optimization geometries is available at \url{https://github.com/choderalab/geometry-benchmark-espaloma}, which is based on the OpenFF Infrastructures \protect\footnote{\url{https://github.com/openforcefield/openff-sage/tree/main/inputs-and-results/benchmarks/qc-opt-geo}} used to validate and assess OpenFF 2.0.0 (Sage)~\cite{sage}.

The QM-optimized conformer geometries and energies utilized in this study were obtained from \texttt{OpenFF Industry Benchmark Season 1 v1.1}
\protect\footnote{\url{https://github.com/openforcefield/qca-dataset-submission/tree/master/submissions/2021-06-04-OpenFF-Industry-Benchmark-Season-1-v1.1}}~\cite{damore2022collaborative} deposited in QCArchive, which was generated at B3LYP-D3BJ/DZVP level of theory.
This dataset consists nearly 9847 unique molecules and 76\,713 conformers of drug-like molecules with mean molecular weight of 348 Da, and a maximum weight of 1104 Da.
It includes formal charges of [-2, -1, 0, 1, 2] and covers atom elements of [Br, F, P, H, N, S, Cl, O, C].
The final benchmarking set consists 9728 unique molecules and 73\,301 conformers, after filtering out connectivity changes during optimization, cases with stereochemistry which cannot be perceived, as well as any calculation failures due to convergence issues.

The QM-optimized molecules were minimized either with \texttt{espaloma-0.3}, \texttt{espaloma-0.3-rc1}, \texttt{openff-2.0.0}, \texttt{openff-2.1.0}, or \texttt{gaff-2.11} force fields using a L-BFGS optimizer implemented in OpenMM 8.0.0~\cite{eastman2017openmm} with a 5.0E-9 kJ/mol/nm convergence tolerance or maximum iteration set to 1500.

The MM-optimized molecules were assessed by measuring the root mean squared deviation (RMSD) in geometries between MM- and QM-optimized conformers, torsion fingerprint deviation (TFD), and error in relative conformer energies (ddE or $\Delta\Delta E$).
The heavy atoms were used to superpose the MM- and QM-optimized molecules to compute the RMSD value using OpenEye Toolkits.
TFD is a weighted metric of deviations in dihedral angles which overcomes the limitations of RMSD~\cite{tanja2012tfd}, which was computed using the RDKit package.
$\Delta\Delta E$ is the energy difference between the MM and QM energies of conformer $\mathbf{x}_\mathrm{i}$, each with respect to the QM minimum energy conformer $\mathbf{x}_\mathrm{0,QM}$:

\begin{equation}
\label{eq:ddE}
\Delta\Delta{E}_i = \Delta E_{\mathrm{MM,i}} - \Delta E_{\mathrm{QM,i}}
= [E_{\mathrm{MM}}(\mathbf{x}_i) - E_{\mathrm{MM}}(\mathbf{x}_\mathrm{0,QM})] 
- [E_{\mathrm{QM}}(\mathbf{x}_i) - E_{\mathrm{QM}}(\mathbf{x}_\mathrm{0,QM})] 
\end{equation}

\section{MD simulations of peptides and calculation of NMR scalar couplings}
\label{sec:si_biopolymer}

A total of 121 experimental NMR observables are available for five homopeptides Ala$_3$, Ala$_4$, Ala$_5$, Gly$_3$, and Val$_3$~\cite{graf2007structure} as well as eight 3-mers Gly-X-Gly, where X is Ala, Glu, Phe, Lys, Leu, Met, Ser, or Val~\cite{hagaram2010intrinsic}.
The initial structure for MD simulations was an extended conformation in which all backbone angles are \ang{180}, constructed from the amino acid sequence using the program \texttt{pmx}~\cite{gapsys2015pmx}.
Protonation states were assigned at pH 2, consistent with the pH of the NMR experiments, using the PROPKA algorithm~\cite{sondergaard2011improved} in the program PDB2PQR 3.6.1~\cite{jurrus2018improvements}.
The peptides were then solvated in a rhombic dodecahedron of TIP3P water~\cite{jorgensen1983comparison} with \SI{1.4}{\nano\meter} padding and neutralizing sodium and chloride counterions using the Modeller module in OpenMM 8.0.0~\cite{eastman2017openmm}.
Monovalent ions were modeled using parameters from Joung and Cheatham~\cite{joung2008determination}.
Force field parameters were assigned to the peptides using either Amber ff14SB~\cite{ff14SB} or \texttt{espaloma-0.3.2}, which is equivalent to \texttt{espaloma-0.3}.
For \texttt{ff14SB}, RESP charges for Ala, Gly, and Val residues with protonated C termini were taken from Nerenberg and Head-Gordon~\cite{nerenberg2011optimizing}.
Hydrogen mass repartitioning with a hydrogen mass of 3.0 amu was applied to the solutes.
The solvated systems were energy minimized with Cartesian restraints applied to non-hydrogen solute atoms with an energy constant of \SI{1.0}{\kilo\calorie\per\mole\per\square\angstrom}.

MD simulations were performed using the CUDA platform of OpenMM 8.0.0~\cite{eastman2017openmm} with a Langevin Middle Integrator~\cite{zhang2019unified}, a Monte Carlo barostat~\cite{bernetti2020pressure}, and constraints on covalent hydrogen bond lengths.
The barostat equilibrium pressure was \SI{1}{\atmosphere}, and the thermostat equilibrium temperature was \SI{300}{\kelvin} for the five homopeptides and \SI{298}{\kelvin} for the eight Gly-X-Gly 3-mers.
During a \SI{1}{\nano\second} equilibration period, the integrator time step was \SI{1}{\femto\second}, the Langevin collision rate was \SI{5}{\per\pico\second}, and the barostat frequency was 5 steps.
During a \SI{500}{\nano\second} production period, the integrator time step was \SI{4}{\femto\second}, the Langevin collision rate was \SI{1}{\per\pico\second}, and the barostat frequency was 25 steps.
Each peptide system and solute force field was simulated using three replicas.

Peptide backbone dihedral angles were extracted from trajectories using the program LOOS 4.0.4~\cite{romo2014lightweight}.
$^3J_{\mathsf{HN,CA}}$ scalar couplings were estimated using parameters from Hennig et al.~\cite{hennig2000determination}.

\begin{multline}
\label{eq:3j_hn_ca}
^3J_{\mathsf{HN,CA}}(\phi_i, \psi_{i-1}) = -0.23 \cos \phi_i - 0.20 \cos \psi_{i-1} + 0.07 \sin \phi_i + 0.08 \cos \psi_{i-1} \\
+ 0.07 \cos \phi_i \cos \psi_{i-1} + 0.12 \cos \phi_i \sin \psi_{i-1} - 0.08 \sin \phi_i \cos \psi_{i-1} - 0.14 \sin \phi_i \sin \psi_{i-1} + 0.54
\end{multline}
where $\phi_i$ is the $\phi$ dihedral angle for the current residue and $\psi_{i-1}$ is the $\psi$ dihedral angle for the previous residue. For all other scalar couplings, the scalar couplings were estimated using a Karplus model~\cite{karplus1963vicinal}.

\begin{equation}
\label{eq:karplus}
J(\theta) = A \cos^2 (\theta + \Delta) + B \cos (\theta + \Delta) + C
\end{equation}
where $\theta$ is the dihedral angle associated with the observable and $A$, $B$, $C$, and $\Delta$ are empirical Karplus parameters~\cite{wirmer2002angular,ding2004protein,hennig2000determination,vogeli2007limits} summarized in \textbf{Table~\ref{tab:si_karplus}}.

Agreement with experiment was quantitatively assessed by computing $\chi^2$ values

\begin{equation}
\label{eq:chisq}
\chi^2 = \frac {1} {N_{\mathsf{obs}}} \sum_{\mathsf{obs}} \frac {\left( J_{\mathsf{comp}} - J_{\mathsf{exp}} \right)^2} {\sigma_{\mathsf{model}}^2}
\end{equation}
where the summation runs over observables, $J_{\mathsf{comp}}$ is the computed scalar coupling estimated using Eq.~\ref{eq:karplus} or Eq.~\ref{eq:3j_hn_ca} averaged over all replicas, $J_{\mathsf{exp}}$ is the experimentally measured scalar coupling, and $\sigma_{\mathsf{model}}$ is the systematic error in the Karplus model, which is an order of magnitude larger than the uncertainty in the experimental values~\cite{graf2007structure,hagaram2010intrinsic}.
The estimates of the systematic uncertainties in the Karplus models are provided in \textbf{Table~\ref{tab:si_karplus}}.

All code used to setup, run, and analyze the peptide MD simulations---including experimental observables and model parameters---can be found at \url{https://github.com/openforcefield/proteinbenchmark}.

\section{Espaloma refitting experiment}
\label{sec:si_refit}
The Python code used to refit and evaluate \texttt{espaloma-0.3} is available at \url{https://github.com/choderalab/refit-espaloma}.
It should be noted that \texttt{espaloma-0.3} is no longer compatible with \texttt{espaloma-0.2.x} models and vice versa.

\subsection{Data preparation}
The quantum chemical datasets obtained from the QCArchive~\cite{qcarchive} in {\bf SI Section~\ref{sec:si_dataset}} were preprocessed prior to the refitting experiment. Molecules with a gap between the minimum and maximum energy larger than 0.1~Hartree (62.5~kcal/mol) were excluded. Since the van der Waals parameters affect the physical property prediction, which is computationally challenging to optimize, we focus on optimizing the valence parameters and use \texttt{openff-2.0.0} force field~\cite{sage} for the van der Waals parameters. AM1-BCC~\cite{jakalian2000fast,jakalian2002fast} ELF10~\footnote{
ELF10 denotes that the ELF ("electrostatically least-interacting functional groups") conformer selection process was used to generate 10 diverse conformations from the lowest energy 2\% of conformers. 
Electrostatic energies are assessed by computing the sum of all Coulomb interactions in vacuum using the absolute values of MMFF charges assigned to each atom~\cite{mmff}. 
AM1-BCC charges~\cite{jakalian2000fast,jakalian2002fast} are generated for each conformer and then averaged.} partial charges were pre-computed using the OpenEye Toolkits as reference charges. These charges were then used to predict the atomic partial charges based on the predicted electronegativity and hardness of atoms, following the same protocol described in the earlier works by Wang et al.~\cite{D2SC02739A}. To ensure that each molecule was represented only once, duplicate molecules across different datasets were merged, ensuring that unique molecules were distributed among the train, validate, or test dataset.

\subsection{Machine learning experimental details}
\subsubsection{Input features}
One of the improvements made from the previous Espaloma framework~\cite{D2SC02739A} is the exclusion of resonance-sensitive features, such as valences and formal charges, in order to improve the handling of molecules with atomic resonance, such as guanidinium and carboxylic acid. In this study, the input features of the atoms included the one-hot encoded element, as well as the hybridization, aromaticity, ring membership of sizes 3 to 8, atom mass, and the degree of the atoms, which is defined as the  number of directly-bonded neighbors, all assigned using the RDKit 2023-03-4 release package \cite{rdkit-2023-03-4}.

\subsubsection{Data splitting and augmentation}
To handle molecular graphs with varying numbers of conformers, all molecules were divided into sets of 50 conformers during training. If there were fewer than 50 conformers, additional ones were randomly selected to reach a total of 50 conformers.
This enabled mini-batching with randomized molecules, making the training process more stochastic compared to the previous study~\cite{D2SC02739A}, where the mini-batch was applied to set of molecules with the same number of conformers rather than individual molecules.

\subsubsection{Hyperparameter optimization}
The hyperparameters were briefly optimized utilizing a subset of data from {\bf SI Section~\ref{sec:si_dataset}}, which included \texttt{OpenFF Gen2-Opt}, \texttt{SPICE-Dipeptide}, and \texttt{RNA-Diverse} datasets. The data was partitioned into train : validate : test sets in a 40:30:30 ratio. During the training process, energy and force matching were applied, along with partial charge fitting using the charge equilibrium approach~\cite{D2SC02739A, wang2023espalomacharge}.


\begin{equation}
\mathcal{L} = W_\mathtt{energy}\mathcal{L}_\mathtt{energy}
+ W_\mathtt{force}\mathcal{L}_\mathtt{force}
+ W_\mathtt{charge}\mathcal{L}_\mathtt{charge}
\end{equation}

Following the protocol specified in Wang et al.~\cite{D2SC02739A}, we utilized GraphSAGE \cite{hamilton2017inductive} as the graph neural network model, the Adam optimizer~\cite{kingma2014adam}, and the Rectified Linear Unit (ReLU) activation function, while maintaining the energy and charge loss weights to 1 and 1e-3, respectively, throughout the optimization experiment. The hyperparameters subject to optimization included the batch size (32, 64, 128, 256), the depth of the graph neural network (2, 3, 4, 5), the depth of the Janossy pooling network (2, 3, 4, 5), the learning rates (1e-3, 1e-4, 5e-5, 1e-5), the number of units per layer (64, 128, 256, 512), and the force weights (1, 1e-1, 1e-2, 1e-3, 1e-4) via grid search on the validation set, and trained for 3000 epochs for each optimization experiment. 

As a result, the optimal configuration was determined as follows: For the atom embedding stage ({\bf Stage1}), three GraphSAGE layers with 512 units and ReLU activation function were employed. For the symmetry preserving pooling stage ({\bf Stage2}) and the readout stage ({\bf Stage3}), four feed-forward layers with 512 units and ReLU activation, a learning rate of 1e-4, and a force loss weight of 1.

\subsubsection{Production run}
The datasets from {\bf SI Section~\ref{sec:si_dataset}} were partitioned into train, validate, and test sets with a distribution of 80:10:10 ratio, respectively, with few exceptions. Notably, the entire \texttt{RNA-Nucleoside} dataset was exclusively utilized for the train set, while the entire \texttt{RNA-Trinucleoside} dataset was allocated for the test set. This partitioning scheme was designed to incorporate diverse molecular structures and enable a comprehensive evaluation of the performance of the espaloma model.

\begin{figure}[tb]
    \centering
    \includegraphics[width=0.8\textwidth]{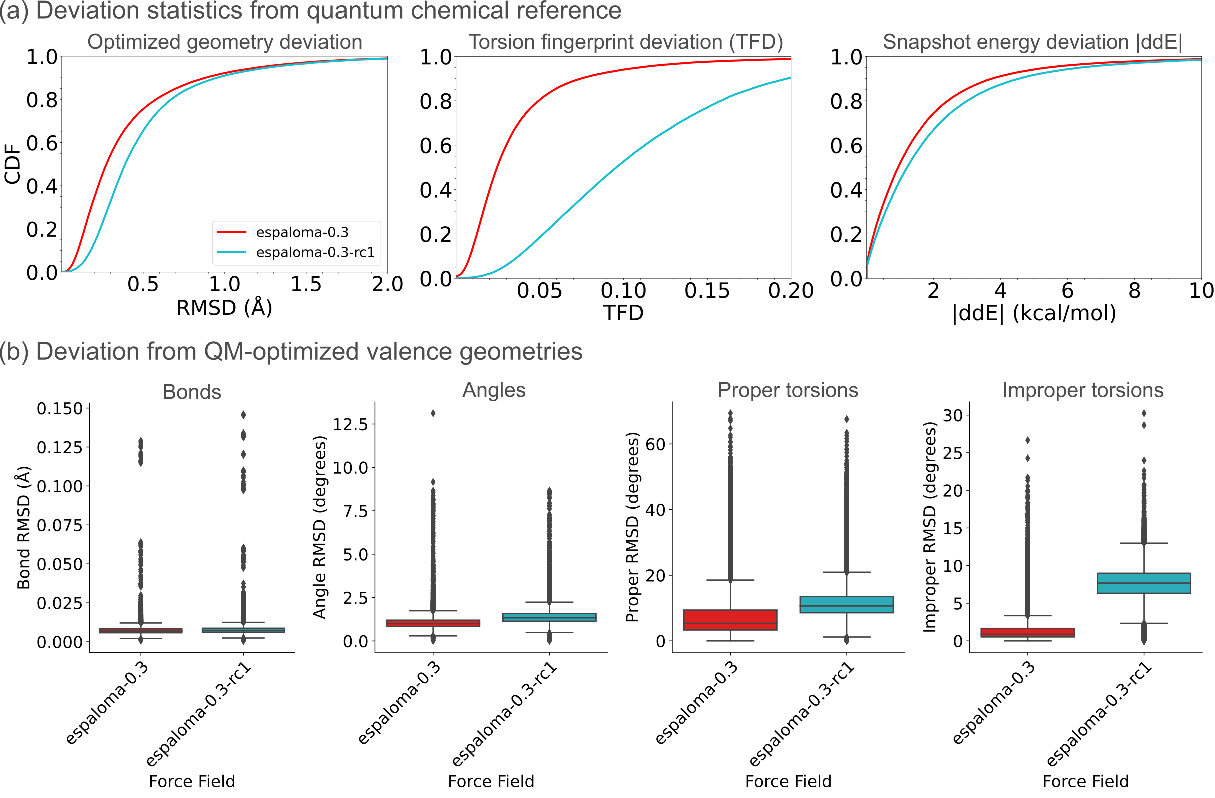}
    \caption{\label{fig:offopt_si}
    \textbf{Espaloma trained with regularizations against torsion terms can better preserve quantum chemical energy minima.} 
    A benchmark of gas-phase QM-optimized geometries, namely \texttt{OpenFF Industry Benchmark Season 1 v1.1}~\cite{damore2022collaborative} from QCArchive, comprising nearly 9728 unique molecules and 73\,301 conformers, was used to compare the structures and energetics of conformers optimized with \texttt{espaloma-0.3} and \texttt{espaloma-0.3-rc1} with respect to their QM-optimized goemetries at the B3LYP-D3BJ/DZVP level of theory. 
    \texttt{espaloma-0.3-rc1} is a model created using the hyperparameters determined during its tuning process (see {\bf Section~\ref{sec:si_refit}}), which does not apply any regularizations to torsion terms.
    (a) The cumulative distribution functions of root-mean-square deviation of atomic positions (RMSD), torsion fingerprint deviation (TFD) score, and relative energy differences (ddE) as described in a previous work~\cite{lim2020f1000} are reported.
    (b) Distributions of bond, angle, proper torsion, and improper torsion RMSD within each conformer with respect to its QM-optimized geometries are shown as quartile box plots. Lower values for all metrics indicate that the MM-optimized geometry is close to the quantum chemical reference structure.
    }
\end{figure}

It should be noted that the espaloma model (\texttt{espaloma-0.3-rc1}), trained with the hyperparameters described above, reproduced torsion profiles poorly compared to its quantum chemical reference structures ({\bf SI Figure{~\ref{fig:offopt_si}}}).
We found that this problem could be remedied by truncating the improper torsion terms to only $n=1, 2$ periodicities, instead of $n=1,...,6$ as in the original method~\cite{D2SC02739A}, and by utilizing regularization for the proper and improper torsion force constants.
Regarding these findings, the final espaloma model was trained with the following loss function with all weights set to 1:

\begin{equation}
\mathcal{L} = W_\mathtt{energy} * \mathcal{L}_\mathtt{energy}
+ W_\mathtt{force} * \mathcal{L}_\mathtt{force}
+ W_\mathtt{proper} * \mathcal{L}_\mathtt{proper}
+ W_\mathtt{improper} * \mathcal{L}_\mathtt{improper}
\end{equation}

To prevent overfitting and ensure optimal model performance, we applied dropouts to the atom embedding stage ({\bf Stage 1}) and symmetry-preserving stage ({\bf Stage 2}), as well as implemented an early stopping mechanism. 
After 800 epochs, the joint root mean square error (RMSE) loss, which incorporates both energies and forces, was monitored using the validation set. This approach allowed us to identify the point at which further training no longer improved the model's generalization capability.

\section{Protein-ligand benchmark dataset}
\label{sec:si_plbenchmark}
The protein-ligand benchmark dataset can be found at \url{https://github.com/kntkb/protein-ligand-benchmark-custom}. 
It consists of 4 target systems (Tyk2, Cdk2, P38, and Mcl1) and a total of 76 ligands. 
This dataset was curated from the openforcefield/protein-ligand-benchmark repository (\url{https://github.com/openforcefield/protein-ligand-benchmark/tree/d3387602bbeb0167abf00dfb81753d8936775dd2}).
Note that one of the ligand from P38 (\texttt{ligand\_p38a\_2ff}) was excluded from the dataset because of its ambigous stereochemistry.
The protein structures and ligand poses, as well as the ligand transformations, were manually curated, while the experimental results were adopted from the original repository. The protein and ligand structures were prepared using Maestro from Schrodinger 2022-2. 

The PDB structure of a protein-ligand complex was imported and processed using the default settings of \texttt{prepwizard}, along with additional options including filling in missing side chains and loops using Prime, capping termini, and deleting waters beyond 5.0~{\AA} from het groups. The  tautomer states of the ligand complexed with the protein were manually inspected, and the most reasonable state was chosen from a human perspective. For the protein residues, the protonation and tautomer states were optimized using the default settings of \texttt{H-bond assignment}. Subsequently, a restrained minimization was performed using the OPLS4 force field, with an RMSD convergence threshold of 0.3~{\AA} for the heavy atoms. The minimized  protein structure from the complex served as the initial protein structure, and X-ray water molecules were retained if necessary, such as buried water molecules in the binding pocket.

For the ligand poses, a flexible ligand alignment approach was applied with respect to the PDB ligand pose found in the protein-ligand complex structure. The default settings of \texttt{ligprep} were used to generate all possible ligand tautomer states, which were then visually inspected to choose the most reasonable state. Subsequently, ligand alignment was performed by aligning all ligands to the PDB ligand pose found in the protein-ligand complex structure, using the \texttt{Ligand Alignment} module in Maestro with Bemis-Murcko scaffold or maximum common scaffold constrain. The ligand poses were manually adjusted, taking into account the binding site environment, which involved rotating ligand torsions and minimizing selected atoms to alleviate severe atom clashes and obtain better initial poses.

Finally, the ligand transformation networks were defined manually by human experts, creating a outward radial map with the simplest ligand in the center. In the case of P38 and Mcl1, R-group substituent from multiple scaffold positions and scaffold hopping were observed. In such cases, ligand transformations were grouped into categories to resemble different structure-activity relationship purposes while maintaining a simplified ligand transformation network.
\section{Alchemical free energy calculations using protein-ligand benchmark dataset}
\label{sec:si_perses}
The Python code used to perform the alchemical protein-ligand binding free energy benchmark experiment is available at \url{https://github.com/choderalab/pl-benchmark-espaloma-experiment}. We utilized the Perses 0.10.1 relative alchemical free energy calculation infrastructure~\cite{perses-0.10.1}, which is based on OpenMM 8.0.0~\cite{eastman2017openmm, eastman2023openmm}, openmmtools 0.22.1~\cite{openmmtools-0.22.1}, and a modified version of openmmforcefields 0.11.0 package~\cite{openmmforcefields} (\url{https://github.com/kntkb/openmmforcefields/tree/6d2c3dcd33d9800a32032d28b6b2dca92f348a43}) to support \texttt{espaloam-0.3}.

All systems were solvated with TIP3P water~\cite{jorgensen1983comparison} with 9.0 Å buffer around the protein, and the system was neutralized with the Joung and Cheatham monovalent counterions~\cite{joung2008determination} with 300 mM NaCl salt concentration. The protein was parametrized with Amber ff14SB force field~\cite{ff14SB}, and the small molecules were parametrized with \texttt{openff-2.1.0}~\cite{openff-2.1.0}, \texttt{espaloma-0.3}, or \texttt{espaloma-0.2.2}~\cite{D2SC02739A}. Additionally, the protein-ligand was self-consistently parametrized with \texttt{espaloam-0.3}, and a modified version of Perses 0.10.1 (\url{https://github.com/kntkb/perses/tree/0d069fc1cf31b8cce1ae7a1482c3fa46bc1382d2}) was used to perform the protein-ligand binding free energy calculations.

Alchemical free energy calculations were simulated with replica exchange among Hamiltonians with Gibbs sampling \cite{chodera2011replica}. All simulations were performed with 12 alchemical states for 10 ns/replica for Tyk2 and Cdk2, 15 ns/replica for Mcl1, and 20 ns/replica for P38, with replica exchange attempts made every 1 ps. The simulations were performed at 300 K and 1 atm using a Monte Carlo Barostat~\cite{bernetti2020pressure} and Langevin BAOAB integrator \cite{leimkuhler2016efficient} with a collision rate of 1/ps. Bonds to hydrogen were constrained, and hydrogen atom masses were set to 3.0 amu by transferring the masses connected to the heavy atoms, allowing for simulations with a 4 fs timestep.

Atom mappings were generated from the provided geometries in the curated benchmark set (see {\bf SI Section~\ref{sec:si_plbenchmark}}). Atoms within 0.5~{\AA} of the transforming ligand pairs were detected as valid mapping atoms using the \texttt{use\_given\_geometries} functionality in Perses.

PyMBAR 3.1.1~\cite{shirts2008statistically} was used to compute the relative free energy, while absolute free energies up to an additive constant were estimated using a least-squares estimation strategy \cite{xu2019optimal} using a modified version of OpenFE cinnabar 0.3.0 package \cite{arsenic} (\url{https://github.com/kntkb/cinnabar/tree/de7bc6623fb25d75848aa1c9f538b77cd02a4b01}). Both experimental and calculated absolute free energies were shifted to their respective means before computing the statistics.

\section{Tyk2 protein-ligand complex MD simulations}
\label{sec:si_vanilla}
\begin{figure}[tb]
    \centering
    \includegraphics[width=0.9\textwidth]{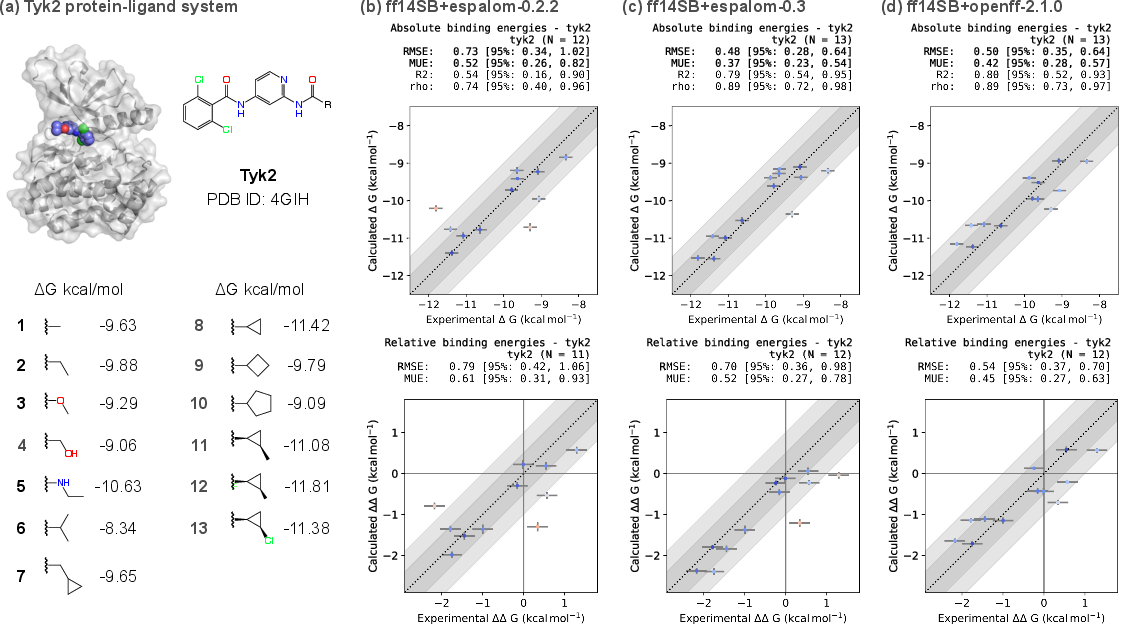}
    \caption{\label{fig:tyk2}
    \textbf{Training \texttt{espaloma-0.3} on an expanded quantum chemical dataset improves protein-ligand binding affinity on the Tyk2 system.}
    (a) We show the X-ray structure used for free energy calculation, along with the 2D structures of all ligands in the Tyk2 protein-ligand benchmark dataset. An outward radial map with ligand \#1 in the center was used for the alchemical ligand transformations.
    We used the Perses 0.10.1 relative free energy calculation infrastructure~\cite{perses-0.10.1} to calculate the relative free energy and assess the performance of (b) \texttt{espaloma-0.2.2}~\cite{D2SC02739A}, (c) \texttt{espaloma-0.3}, and (d) \texttt{openff-2.1.0}~\cite{openff-2.1.0} by parametrizing the small molecules with each force field.
    Amber ff14SB force field~\cite{ff14SB} was used to parametrize the protein for all cases. Notably, \texttt{espaloma-0.2.2} failed to simulate the alchemical ligand transformation of ligand \#1 to ligand \#2; hence one ligand is not reported in (b). \texttt{espaloma-0.2.2} and \texttt{espaloma-0.3} achieves an absolute free energy ($\Delta G$) RMSE of 0.73 [95\% CI: 0.34, 1.02] kcal/mol and 0.48 [95\% CI: 0.28, 0.64] kcal/mol, respectively, suggesting that \texttt{espaloma-0.3} tends to show improved performance over \texttt{espaloma-0.2.2}.
    }
\end{figure}

The unbiased MD simulation code used in this study, along with initial prepared structures, can be found at \url{https://github.com/choderalab/vanilla-espaloma-experiment}.
The initial structures of Tyk2 and ligand \#1 shown in {\bf SI Figure{\ref{fig:tyk2}}} was taken from the protein-ligand benchmark dataset as described in {\bf SI Section~\ref{sec:si_plbenchmark}}. The ligand was parametrized with either \texttt{openff-2.1.0}~\cite{openff-2.1.0} or \texttt{espaloma-0.3}, and protein parametrized with Amber ff14SB~\cite{ff14SB}.
The protein-ligand complex system was solvated with TIP3P water~\cite{jorgensen1983comparison} and neutralized with the Joung and Cheatham monovalent counterions~\cite{joung2008determination} with 150 mM NaCl salt concentration.

All simulations were performed at 300 K and 1 atm using a Monte Carlo Barostat~\cite{bernetti2020pressure} and Langevin Middle Integrator (a variant splitting of the BAOAB integrator)~\cite{zhang2019unified} with a collision rate of 1/ps. Bonds to hydrogen were constrained, and hydrogen atom masses were set to 3.0 amu allowing for simulations with a 4 fs timestep. The solvated systems were minimized and subsequently subjected to 1 microsecond of simulation using OpenMM 8.0.0~\cite{eastman2017openmm}.

The root-mean square deviation (RMSD) profile of the heavy ligand atoms and protein C${\alpha}$ atoms were reported over the 1 microsecond MD simulation. The trajectories were aligned with respect to the binding pocket residues (within 4 ${\text{Å}}$ from the initial ligand pose) before computing the heavy ligand atom RMSD. Similarly, the protein C${\alpha}$ atoms excluding the first and last 5 residues, were used to align the protein trajectories before RMSD calculation, with the first and last 5 residues excluded from RMSD computation.

\begin{table}[tbp]
\centering
\resizebox{\columnwidth}{!}{%
\begin{tabular}{ccccccccccc}
\toprule
  {\begin{tabular}[c]{c@{}}
      Dataset\\
      {\small (QCArchive Workflow)}
  \end{tabular}} &  
  Category &
  Mols &
  Confs &
  Split &
  \multicolumn{3}{c}
    {\begin{tabular}[c]{@{}c@{}}
        {\bf \large espaloma-0.3}\\ 
        {\footnotesize Energy RMSE (kcal/mol)}\\ 
        {\footnotesize Force RMSE (kcal/mol $\cdot$ ${\text{Å}}^{-1}$)}
    \end{tabular}} &
  \multicolumn{3}{c}
    {\begin{tabular}[c]{@{}c@{}}{\bf \large Repetition}\\ 
    {\footnotesize Energy RMSE (kcal/mol)}\\ 
    {\footnotesize Force RMSE (kcal/mol $\cdot$ ${\text{Å}}^{-1}$)}
    \end{tabular}} \\
  \cmidrule(r){6-8} \cmidrule(r){9-11}
   &
   &
   &
   &
   &
  Train (80\%) &
  Validate (10\%) &
  Test (10\%) &
  Train (80\%) &
  Validate (10\%) &
  Test (10\%) \\
  \midrule
  \begin{tabular}[c]{@{}c@{}}{\bf SPICE-Pubchem~\cite{eastman2023spice, kim2023pubchem}}\\{\small (Dataset)}\end{tabular} &
  \begin{tabular}[c]{@{}c@{}}Small\\ molecule\end{tabular} &
  14110 &
  608436 &
  {\footnotesize 80:10:10} &
  \begin{tabular}[c]{@{}c@{}}
    $2.06_{2.04}^{2.07}$\\$6.22_{6.19}^{6.26}$
  \end{tabular} &
  \begin{tabular}[c]{@{}c@{}}
    $2.31_{2.25}^{2.37}$\\$6.79_{6.65}^{6.95}$
  \end{tabular} &
  \begin{tabular}[c]{@{}c@{}}
    $2.30_{2.25}^{2.36}$\\$6.81_{6.68}^{6.95}$
  \end{tabular} &
  \begin{tabular}[c]{@{}c@{}}
    $2.01_{1.99}^{2.03}$\\$6.18_{6.15}^{6.21}$
  \end{tabular} &
  \begin{tabular}[c]{@{}c@{}}
    $2.23_{2.19}^{2.28}$\\$6.73_{6.57}^{6.94}$
  \end{tabular} &
  \begin{tabular}[c]{@{}c@{}}
    $2.25_{2.20}^{2.30}$\\$6.64_{6.51}^{6.78}$
  \end{tabular} 
  \vspace{1mm} \\

  \begin{tabular}[c]{@{}c@{}}{\bf SPICE-DES-Monomers~\cite{eastman2023spice, DES370K}}\\{\small (Dataset)}\end{tabular} &
  \begin{tabular}[c]{@{}c@{}}Small\\ molecule\end{tabular} &
  369 &
  18435 &
  {\footnotesize 80:10:10} &
  \begin{tabular}[c]{@{}c@{}}
    $1.39_{1.32}^{1.46}$\\$5.86_{5.69}^{6.02}$
  \end{tabular} &
  \begin{tabular}[c]{@{}c@{}}
    $1.34_{1.13}^{1.60}$\\$5.63_{5.12}^{6.24}$
  \end{tabular} &
  \begin{tabular}[c]{@{}c@{}}
    $1.36_{1.13}^{1.67}$\\$5.91_{5.49}^{6.42}$
  \end{tabular} &
  \begin{tabular}[c]{@{}c@{}}
    $1.36_{1.29}^{1.43}$\\$5.83_{5.66}^{5.99}$
  \end{tabular} &
  \begin{tabular}[c]{@{}c@{}}
    $1.38_{1.13}^{1.68}$\\$5.56_{5.24}^{5.96}$
  \end{tabular} &
  \begin{tabular}[c]{@{}c@{}}
    $1.41_{1.20}^{1.64}$\\$5.92_{5.42}^{6.57}$
  \end{tabular} 
  \vspace{1mm} \\

  \begin{tabular}[c]{@{}c@{}}{\bf Gen2-Opt}\\{\small 
  (OptimizationDataset)}\end{tabular} &
  \begin{tabular}[c]{@{}c@{}}Small\\ molecule\end{tabular} &
  1024 &
  244989 &
  {\footnotesize 80:10:10} &
  \begin{tabular}[c]{@{}c@{}}
    $1.36_{1.26}^{1.48}$\\$3.94_{3.79}^{4.11}$
  \end{tabular} &
  \begin{tabular}[c]{@{}c@{}}
    $1.35_{1.17}^{1.56}$\\$4.22_{3.92}^{4.52}$
  \end{tabular} &
  \begin{tabular}[c]{@{}c@{}}
    $1.66_{1.21}^{2.29}$\\$4.47_{3.90}^{5.40}$
  \end{tabular} &
  \begin{tabular}[c]{@{}c@{}}
    $1.31_{1.20}^{1.43}$\\$3.77_{3.64}^{3.92}$
  \end{tabular} &
  \begin{tabular}[c]{@{}c@{}}
    $1.51_{1.15}^{1.93}$\\$4.76_{3.91}^{6.01}$
  \end{tabular} &
  \begin{tabular}[c]{@{}c@{}}
    $1.41_{1.16}^{1.71}$\\$4.32_{3.71}^{5.09}$
  \end{tabular} 
  \vspace{1mm} \\

  \begin{tabular}[c]{@{}c@{}}{\bf Gen2-Torsion}\\{\small 
  (TorsionDriveDataset)}\end{tabular} &
  \begin{tabular}[c]{@{}c@{}}Small\\ molecule\end{tabular} &
  729 &
  25832 &
  {\footnotesize 80:10:10} &
  \begin{tabular}[c]{@{}c@{}}
    $1.76_{1.61}^{1.91}$\\$4.31_{4.18}^{4.44}$
  \end{tabular} &
  \begin{tabular}[c]{@{}c@{}}
    $1.97_{1.60}^{2.42}$\\$5.00_{4.49}^{5.55}$
  \end{tabular} &
  \begin{tabular}[c]{@{}c@{}}
    $1.64_{1.32}^{2.01}$\\$4.71_{4.18}^{5.29}$
  \end{tabular} &
  \begin{tabular}[c]{@{}c@{}}
    $1.66_{1.52}^{1.79}$\\$4.25_{4.12}^{4.38}$
  \end{tabular} &
  \begin{tabular}[c]{@{}c@{}}
    $1.91_{1.48}^{2.37}$\\$4.56_{4.12}^{5.01}$
  \end{tabular} &
  \begin{tabular}[c]{@{}c@{}}
    $1.84_{1.43}^{2.26}$\\$5.40_{4.26}^{7.03}$
  \end{tabular} 
  \vspace{1mm} \\

  \begin{tabular}[c]{@{}c@{}}{\bf SPICE-Dipeptide~\cite{eastman2023spice}}\\{\small (Dataset)}\end{tabular} &
  Peptide &
  677 &
  26279 &
  {\footnotesize 80:10:10} &
  \begin{tabular}[c]{@{}c@{}}
    $3.21_{3.16}^{3.26}$\\$7.98_{7.88}^{8.07}$
  \end{tabular} &
  \begin{tabular}[c]{@{}c@{}}
    $3.15_{3.01}^{3.30}$\\$8.05_{7.77}^{8.34}$
  \end{tabular} &
  \begin{tabular}[c]{@{}c@{}}
    $3.09_{2.96}^{3.21}$\\$7.78_{7.55}^{8.02}$
  \end{tabular} &
  \begin{tabular}[c]{@{}c@{}}
    $3.06_{3.01}^{3.11}$\\$7.81_{7.71}^{7.90}$
  \end{tabular} &
  \begin{tabular}[c]{@{}c@{}}
    $3.15_{3.02}^{3.29}$\\$7.74_{7.47}^{7.97}$
  \end{tabular} &
  \begin{tabular}[c]{@{}c@{}}
    $2.94_{2.82}^{3.07}$\\$7.64_{7.39}^{7.87}$
  \end{tabular} 
  \vspace{1mm} \\

  \begin{tabular}[c]{@{}c@{}}{\bf Pepconf-Opt~\cite{prasad2019pepconf}}\\{\small 
  (OptimizationDataset)}\end{tabular} &
  Peptide &
  557 &
  166291 &
  {\footnotesize 80:10:10} &
  \begin{tabular}[c]{@{}c@{}}
    $2.61_{2.43}^{2.83}$\\$3.83_{3.60}^{4.09}$
  \end{tabular} &
  \begin{tabular}[c]{@{}c@{}}
    $2.82_{2.41}^{3.27}$\\$3.65_{3.29}^{4.12}$
  \end{tabular} &
  \begin{tabular}[c]{@{}c@{}}
    $2.79_{2.45}^{3.13}$\\$4.01_{3.63}^{4.46}$
  \end{tabular} &
  \begin{tabular}[c]{@{}c@{}}
    $2.56_{2.40}^{2.73}$\\$3.78_{3.58}^{4.02}$
  \end{tabular} &
  \begin{tabular}[c]{@{}c@{}}
    $2.87_{2.24}^{3.77}$\\$3.92_{3.43}^{4.62}$
  \end{tabular} &
  \begin{tabular}[c]{@{}c@{}}
    $3.20_{2.45}^{4.17}$\\$4.29_{3.53}^{5.49}$
  \end{tabular} 
  \vspace{1mm} \\

  \begin{tabular}[c]{@{}c@{}}{\bf Protein-Torsion}\\{\small (TorsionDriveDataset)}\end{tabular} &
  Peptide &
  62 &
  48999 &
  {\footnotesize 80:10:10} &
  \begin{tabular}[c]{@{}c@{}}
    $2.27_{2.06}^{2.50}$\\$3.94_{3.70}^{4.24}$
  \end{tabular} &
  \begin{tabular}[c]{@{}c@{}}
    $1.91_{1.36}^{2.28}$\\$3.49_{2.85}^{3.97}$
  \end{tabular} &
  \begin{tabular}[c]{@{}c@{}}
    $1.93_{1.73}^{2.14}$\\$3.49_{3.22}^{3.78}$
  \end{tabular} &
  \begin{tabular}[c]{@{}c@{}}
    $2.20_{2.02}^{2.39}$\\$3.85_{3.56}^{4.19}$
  \end{tabular} &
  \begin{tabular}[c]{@{}c@{}}
    $2.52_{1.85}^{3.16}$\\$4.21_{3.65}^{5.00}$
  \end{tabular} &
  \begin{tabular}[c]{@{}c@{}}
    $2.46_{1.80}^{3.40}$\\$4.01_{3.55}^{4.62}$
  \end{tabular} 
  \vspace{1mm} \\

  \begin{tabular}[c]{@{}c@{}}{\bf RNA-Diverse}\\{\small (Dataset)}\end{tabular} &
  RNA &
  64 &
  3703 &
  {\footnotesize 80:10:10} &
  \begin{tabular}[c]{@{}c@{}}
    $4.12_{3.95}^{4.31}$\\$4.44_{4.40}^{4.47}$
  \end{tabular} &
  \begin{tabular}[c]{@{}c@{}}
    $4.51_{4.05}^{4.92}$\\$4.54_{4.50}^{4.58}$
  \end{tabular} &
  \begin{tabular}[c]{@{}c@{}}
    $4.17_{3.85}^{4.52}$\\$4.41_{4.29}^{4.51}$
  \end{tabular} &
  \begin{tabular}[c]{@{}c@{}}
    $4.13_{3.95}^{4.29}$\\$4.42_{4.39}^{4.46}$
  \end{tabular} &
  \begin{tabular}[c]{@{}c@{}}
    $4.57_{4.04}^{5.18}$\\$4.54_{4.50}^{4.59}$
  \end{tabular} &
  \begin{tabular}[c]{@{}c@{}}
    $4.12_{3.68}^{4.71}$\\$4.47_{4.39}^{4.54}$
  \end{tabular} 
  \vspace{1mm} \\

  \begin{tabular}[c]{@{}c@{}}{\bf RNA-Trinucleotide}\\{\small (Dataset)}\end{tabular} &
  RNA &
  64 &
  35811 &
  {\footnotesize 0:0:100} &
  \begin{tabular}[c]{@{}c@{}}
    ---\\---
  \end{tabular} &
  \begin{tabular}[c]{@{}c@{}}
    ---\\---
  \end{tabular} &
  \begin{tabular}[c]{@{}c@{}}
    $3.75_{3.59}^{3.94}$\\$4.28_{4.20}^{4.39}$
  \end{tabular} &
  \begin{tabular}[c]{@{}c@{}}
    ---\\---
  \end{tabular} &
  \begin{tabular}[c]{@{}c@{}}
    ---\\---
  \end{tabular} &
  \begin{tabular}[c]{@{}c@{}}
    $3.80_{3.64}^{3.97}$\\$4.27_{4.20}^{4.37}$
  \end{tabular} 
  \vspace{1mm} \\

  \begin{tabular}[c]{@{}c@{}}{\bf RNA-Nucleoside}\\{\small (Dataset)}\end{tabular} &
  RNA &
  4 &
  9542 &
  {\footnotesize 100:0:0} &
  \begin{tabular}[c]{@{}c@{}}
    $1.32_{1.16}^{1.49}$\\$4.17_{3.86}^{4.47}$
  \end{tabular} &
  \begin{tabular}[c]{@{}c@{}}
    ---\\---
  \end{tabular} &
  \begin{tabular}[c]{@{}c@{}}
    ---\\---
  \end{tabular} &
  \begin{tabular}[c]{@{}c@{}}
    $1.26_{1.11}^{1.43}$\\$4.00_{3.67}^{4.33}$
  \end{tabular} &
  \begin{tabular}[c]{@{}c@{}}
    ---\\---
  \end{tabular} &
  \begin{tabular}[c]{@{}c@{}}
    ---\\---
  \end{tabular} \\
\bottomrule
\end{tabular}%
}
\caption{{\bf A repeated Espaloma refitting experiment yields consistent results with \texttt{espaloma-0.3}, capable of accurately fitting quantum chemical energies and forces.}
The Espaloma refitting experiment was conducted using a different random seed to partition the datasets into train, validate, and test sets. The RMSE metrics of energy and forces were analyzed similarly to those of \texttt{espaloma-0.3}. The 95\% confidence intervals, annotated in the results, were calculated by bootstrapping molecule replacement using 1000 replicates.
\vspace{-20 pt}
}
\label{tab:si_rmse}
\end{table}
\begin{table}[tbp]
\centering
\resizebox{\columnwidth}{!}{%
\begin{tabular}{
    c
    c
    S[table-format=3.1]
    S[table-format=2.2]
    S[table-format=2.2]
    S[table-format=2.2]
    S[table-format=1.2]
    c
}
\toprule
    Observable &
    $\theta$ &
    \multicolumn{1}{c}{$\Delta$} &
    \multicolumn{1}{c}{$A$} &
    \multicolumn{1}{c}{$B$} &
    \multicolumn{1}{c}{$C$} &
    \multicolumn{1}{c}{$\sigma_{\mathsf{model}}$} &
    Reference \\
    \midrule
    $^1J_{\mathsf{N,CA}}$ &
    $\psi_i$ &
    0.0 &
    1.70 &
    -0.98 &
    9.51 &
    0.59 &
    Wirmer and Schwalbe~\cite{wirmer2002angular} \\
    $^2J_{\mathsf{N,CA}}$ &
    $\psi_{i-1}$ &
    0.0 &
    -0.66 &
    -1.52 &
    7.85 &
    0.50 &
    Ding and Gronenborn~\cite{ding2004protein} \\
    $^3J_{\mathsf{HA,C'}}$ &
    $\phi_i$ &
    120.0 &
    3.72 &
    -2.18 &
    1.28 &
    0.38$^\mathsf{a}$ &
    Hu and Bax~\cite{hu1997determination} \\
    $^3J_{\mathsf{HN,CB}}$ &
    $\phi_i$ &
    60.0 &
    3.51 &
    -0.53 &
    0.14 &
    0.25 &
    V\"{o}geli et al.~\cite{vogeli2007limits} \\
    $^3J_{\mathsf{HN,C'}}$ &
    $\phi_i$ &
    180.0 &
    4.12 &
    -1.10 &
    0.11 &
    0.31 &
    V\"{o}geli et al.~\cite{vogeli2007limits} \\
    $^3J_{\mathsf{HN,HA}}$ &
    $\phi_i$ &
    -60.0 &
    7.97 &
    -1.26 &
    0.63 &
    0.42 &
    V\"{o}geli et al.~\cite{vogeli2007limits} \\
    $^3J_{\mathsf{HN,CA}}$ &
    $\phi_i$, $\psi_{i-1}$ &
    \multicolumn{4}{c}{Eq.~\ref{eq:3j_hn_ca}} &
    0.10 &
    Hennig et al.~\cite{hennig2000determination} \\
\bottomrule
\end{tabular}%
}
\caption{{\bf Karplus parameters used to estimate NMR scalar couplings.}
Empirical Karplus parameters $\Delta$, $A$, $B$, and $C$ used to estimate scalar couplings via Eq.~\ref{eq:karplus} and systematic errors in Karplus models $\sigma$ used to estimate $\chi^2$ values via Eq.~\ref{eq:chisq}.
$^\mathsf{a}$Systematic error estimate for the $^3J_{\mathsf{HA,C'}}$ Karplus model taken from Wickstrom et al.~\cite{wickstrom2009evaluating}
}
\label{tab:si_karplus}
\end{table}
\begin{figure}[tb]
    \centering
    \includegraphics[width=0.9\textwidth]{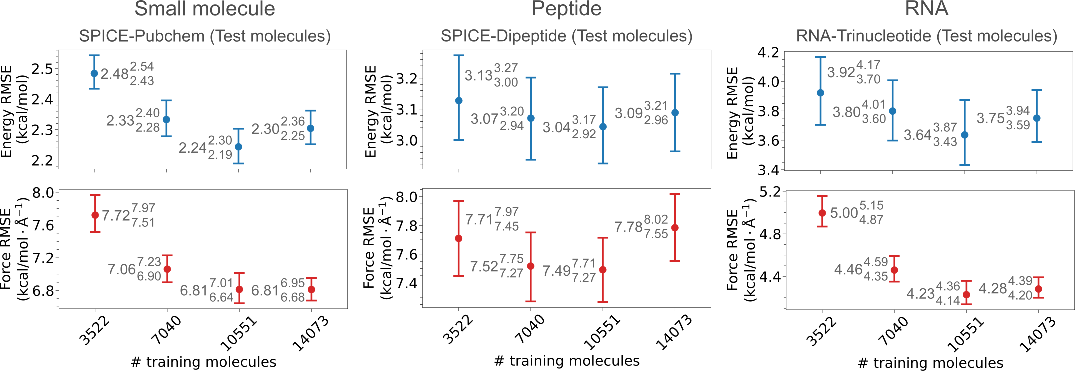}
    \caption{\label{fig:train_size}
    \textbf{Espaloma framework can directly fit to quantum mechanical energies and forces even in low data regimes.}
    The espaloma refitting experiment was conducted with a varying number of molecules in the training set. The same validation and test sets used to develop \texttt{espaloma-0.3} were maintained consistently throughout this experiment. The energy and force RMSE values on the test dataset are reported for the \texttt{SPICE-Pubchem}, \texttt{SPICE-Dipeptide}, and \texttt{RNA-Trinucleotide} datasets to illustrate the outcomes for small molecule, peptide, and RNA chemical series. The 95\% confidence intervals, as annotated in the results, were calculated by bootstrapping molecule replacement using 1000 replicates.
    }
\end{figure}
\begin{figure}[tb]
    \centering
    \includegraphics[width=0.8\textwidth]{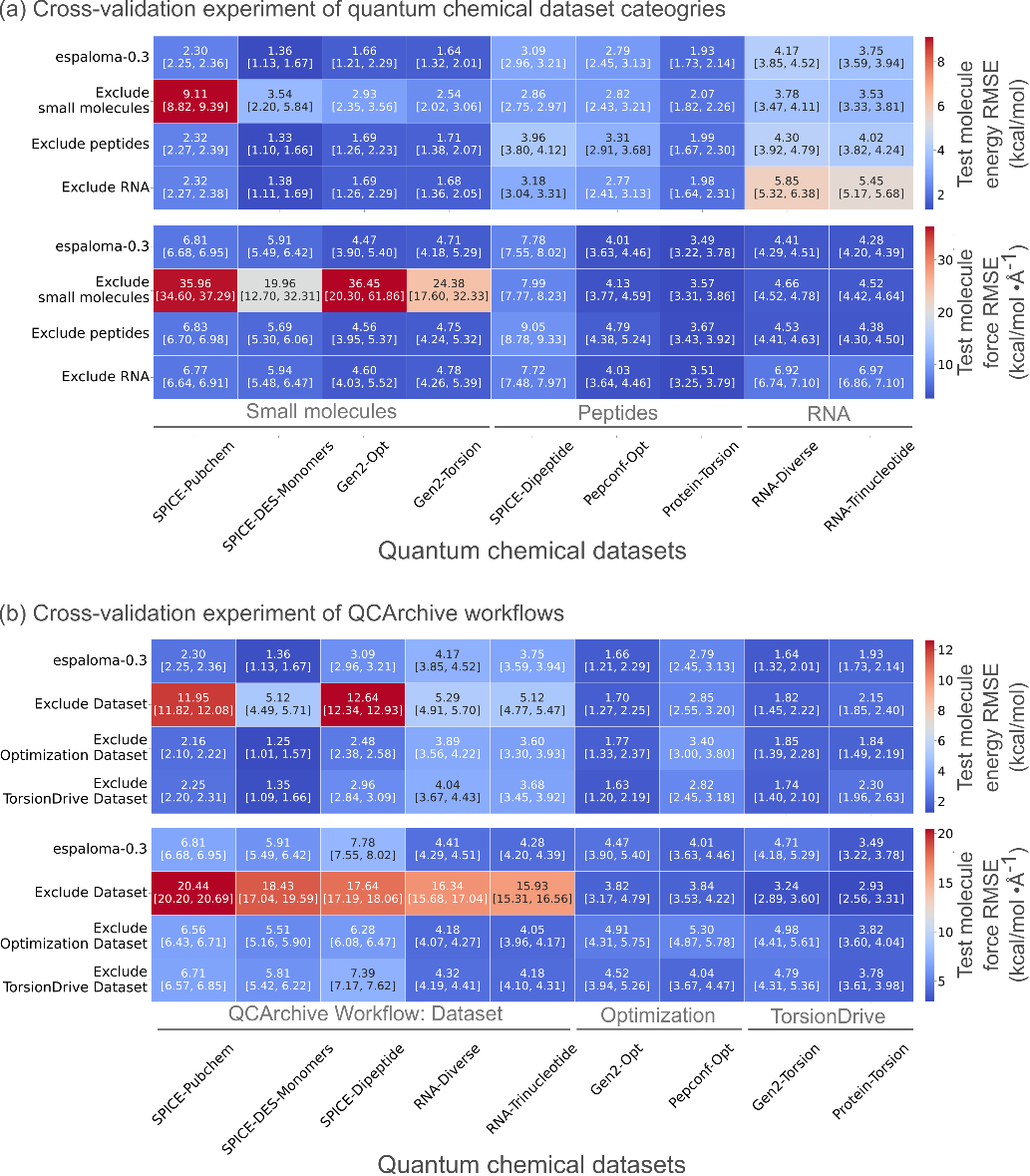}
    \caption{\label{fig:leave_out}
    \textbf{Chemical diversity and high-energy conformers are important for accurately capturing quantum chemical energies and forces with Espaloma.}
    Espaloma refitting experiments were conducted by excluding certain quantum chemical datasets during training and validation, following the procedures outlined in deploying \texttt{espaloma-0.3}. These experiments aimed to investigate how the quantum chemical datasets used for training espaloma affect its ability to accurately reproduce quantum chemical energies and forces. The refitting experiment was conducted with two different scenarios: (a) Quantum chemical datasets corresponding to the small molecules, peptides, or RNA chemical series were excluded from both training and validation; or (b) Quantum chemical datasets generated using the three distinct QCArchive workflows (see {\bf SI Section~\ref{sec:si_dataset}}) --- \texttt{Dataset}, \texttt{Optimization Dataset}, or \texttt{TorsionDrive Dataset} --- were excluded from both training and validation. The energy and force RMSE metrics for the test molecules, including the quantum chemical datasets excluded during training and validation, are reported with 95\% confidence intervals. These intervals were calculated by bootstrapping molecule replacement with 1000 replicas and are depicted in square brackets.
    }
\end{figure}
\begin{figure}[tb]
    \centering
    \includegraphics[width=0.8\textwidth]{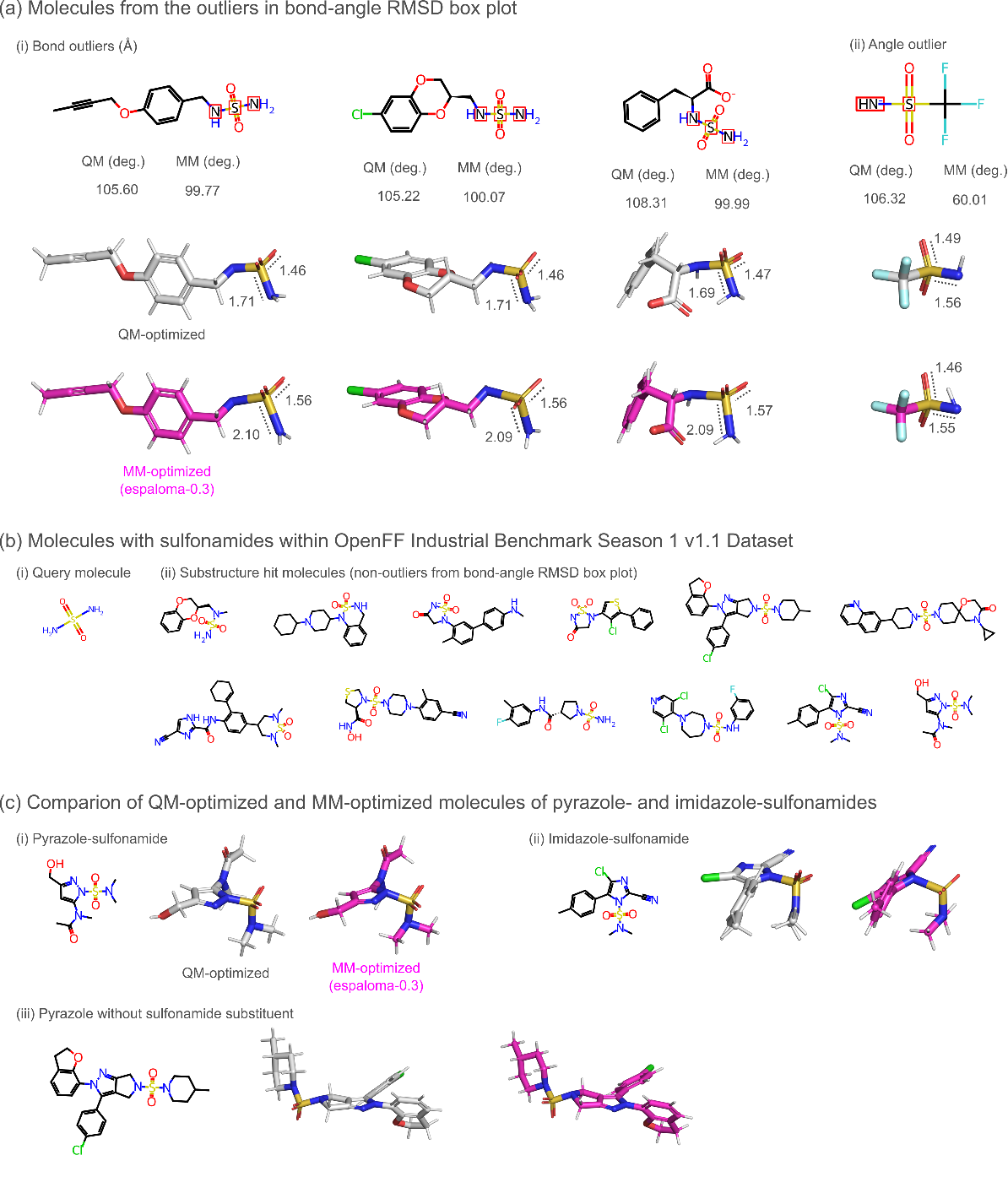}
    \caption{\label{fig:offopt_outlier_si}
    \textbf{Molecules containing sulfonamides are more challenging to maintain their QM-optimized geometries when minimized with espaloma-0.3 compared to other molecules.} 
    (a) Representative molecular conformers identified as outliers in the bond-angle RMSD box plot ({\bf{Figure~\ref{fig:offopt}}}) are shown, where bond RMSD > 0.1~\text{Å} and angle RMSD > 10 degrees were considered as outliers. Three sulfonamide molecules connected to an aliphatic carbon exhibit elongated bond (S-O and S-N) distances respect to QM-optimized geometries, and a single angle outlier with a deviation of $\sim$40 degrees deviation from QM-optimized geometry was observed.
    (b) Molecules containing sulfonamide groups, excluding the outliers in (a) are shown, with each molecular conformer featuring reasonable bond distances within the sulfonamide group.
    (c) The nitrogen geometry of pyrazoles and imidazoles substituted with sulfonamides becomes trigonal pyramidal when minimized with \texttt{espaloma-0.3}, rather than preserving a flat ring geometry and losing their sp2 hybridized features, as observed with QM-optimized geometries.
    }
\end{figure}
\begin{figure}[tb]
    \centering
    \includegraphics[width=0.9\textwidth]{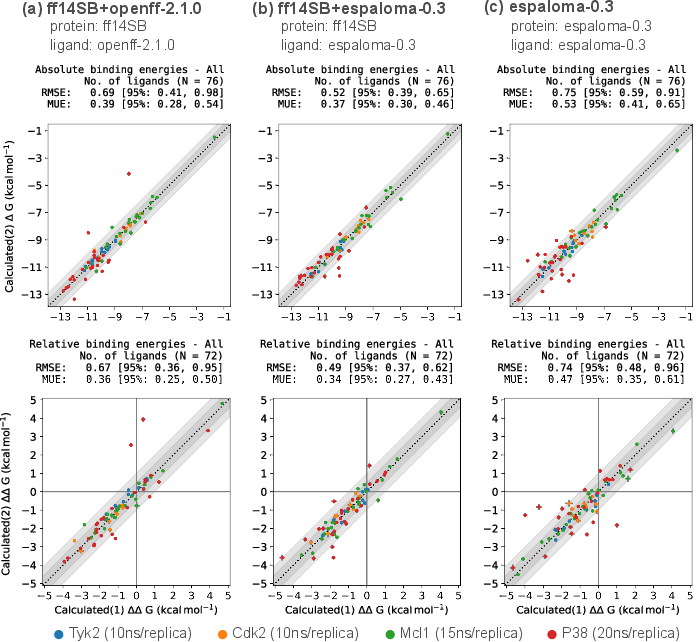}
    \caption{\label{fig:converge}
    \textbf{Alchemical free energy calculations are well-reproduced within 10-20 ns of simulation time.}
    The reproducibility of alchemical protein-ligand free energy calculations described in {\bf Section~\ref{sec:binding}} was investigated by conducting repeated simulations on Tyk2 (10 ns/replica), Cdk2 (10 ns/replica), Mcl1 (15 ns/replica), and P38 (20 ns/replica) using the same simulation protocols. The small molecules were parametrized either with (a) \texttt{openff-2.1.0}, (b) \texttt{espaloma-0.3} combined with Amber ff14SB for proteins, or (c) by paramterizing both small molecule and protein self-consistently with \texttt{espaloma-0.3}. The light and dark gray regions depict the confidence bounds of 0.5 kcal/mol and 1.0 kcal/mol, respectively.
    }
\end{figure}
\begin{figure}[tb]
    \centering
    \includegraphics[width=0.8\textwidth]{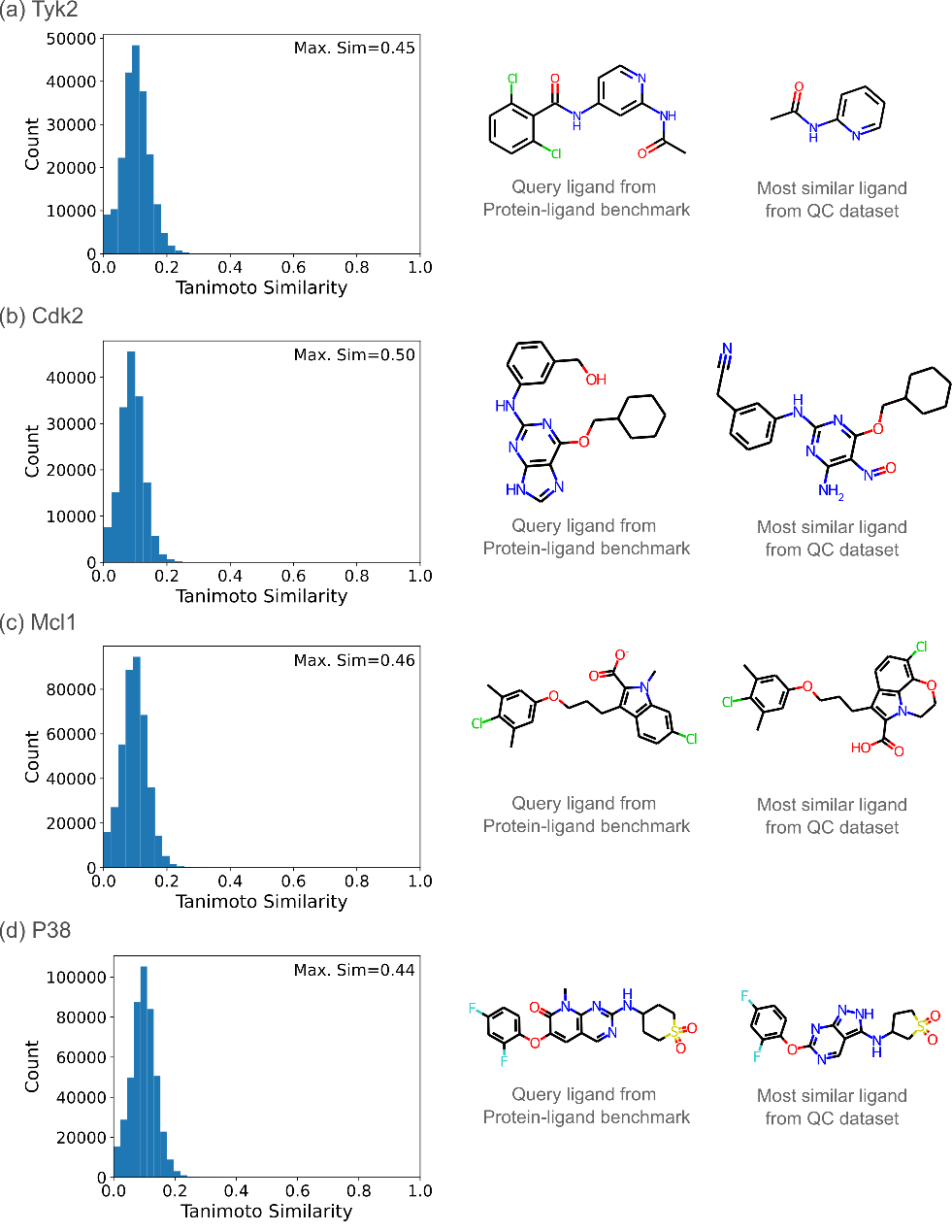}
    \caption{\label{fig:tanimoto}
    \textbf{The ligands from the protein-ligand binding free energy benchmark dataset significantly differ from the quantum chemical (QC) dataset used to train \texttt{espaloma-0.3}.}
    Pairwise Tanimoto similarity scores between the ligands from the protein-ligand binding free energy benchmark dataset and the QC datasets used to deploy \texttt{espaloma-0.3} were investigated. The maximum Tanimoto similarity score is reported for each target system in the protein-ligand binding free energy benchmark dataset, along with the molecular pair that achieved the maximum similarity score.
    }
\end{figure}
\begin{figure}[tb]
    \centering
    \includegraphics[width=0.8\textwidth]{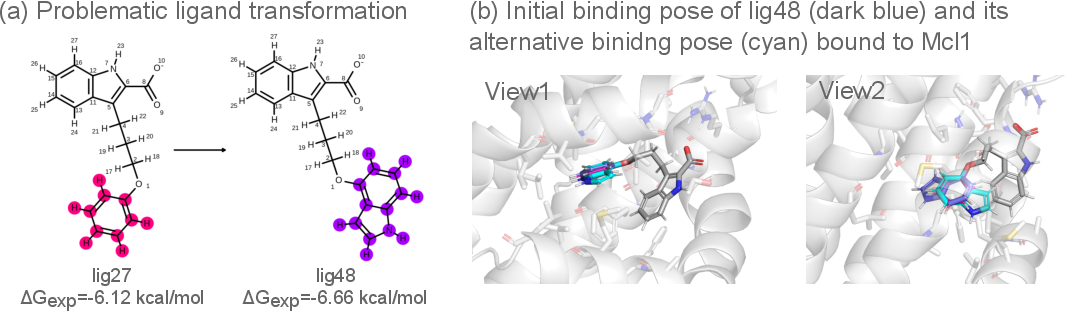}
    \caption{\label{fig:mcl1_outlier}
    \textbf{The alchemical protein-ligand binding free energy calculation for the outlier Mcl1 ligand can be improved by adopting an alternative binding pose.}
    (a) Illustration of the problematic Mcl1 ligand transformation observed as an outlier during the alchemical protein-ligand binding free energy calculation in {\bf Figure~\ref{fig:all}}. The transforming ligand atoms are colored in magenta and purple.
    (b) The initial complex structure of Mcl1, bound with ligand \#48 (in dark blue), used to simulate the alchemical free energy calculations, is illustrated along with its alternative flipped binding pose (in cyan).
    }
\end{figure}
\begin{figure}[tb]
    \centering
    \includegraphics[width=0.9\textwidth]{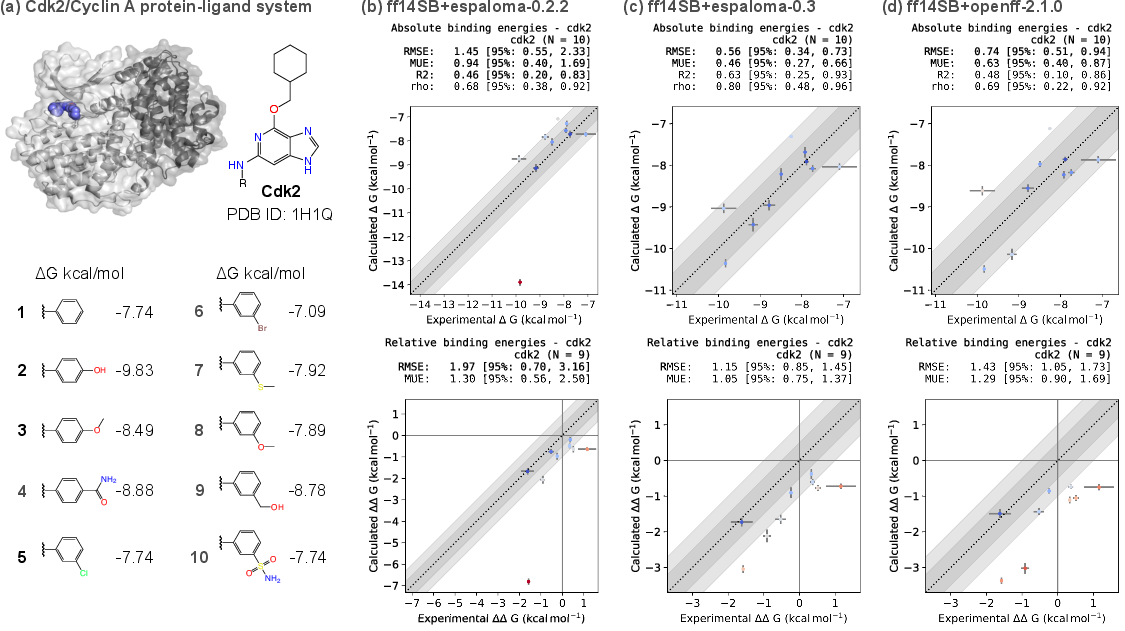}
    \caption{\label{fig:cdk2}
    \textbf{Training \texttt{espaloma-0.3} on an extensive quantum chemical dataset significantly improves protein-ligand binding affinity calculations on the Cdk2 system.}
    (a) We show the X-ray structure used for free energy calculation, along with the 2D structures of all ligands in the Cdk2 protein-ligand benchmark dataset. An outward radial map with ligand \#1 in the center was used for the alchemical ligand transformations.
    We used the Perses 0.10.1 relative free energy calculation infrastructure~\cite{perses-0.10.1} to calculate the relative free energy and assess the performance of (b) \texttt{espaloma-0.2.2}~\cite{D2SC02739A}, (c) \texttt{espaloma-0.3}, and (d) \texttt{openff-2.1.0}~\cite{openff-2.1.0} by parametrizing the small molecules with each force field.
    Amber ff14SB force field~\cite{ff14SB} was used to parametrize the protein for all cases. \texttt{espaloma-0.2.2} and \texttt{espaloma-0.3} achieves an absolute free energy ($\Delta G$) RMSE of 1.45 [95\% CI: 0.55, 2.33] kcal/mol and 0.56 [95\% CI: 0.34, 0.73] kcal/mol, respectively, indicating that \texttt{espaloma-0.3} trained on extensive quantum chemical dataset significantly improved protein-ligand binding affinity calculations on the Cdk2 system.
    }
\end{figure}
\begin{figure}[tb]
    \centering
    \includegraphics[width=0.8\textwidth]{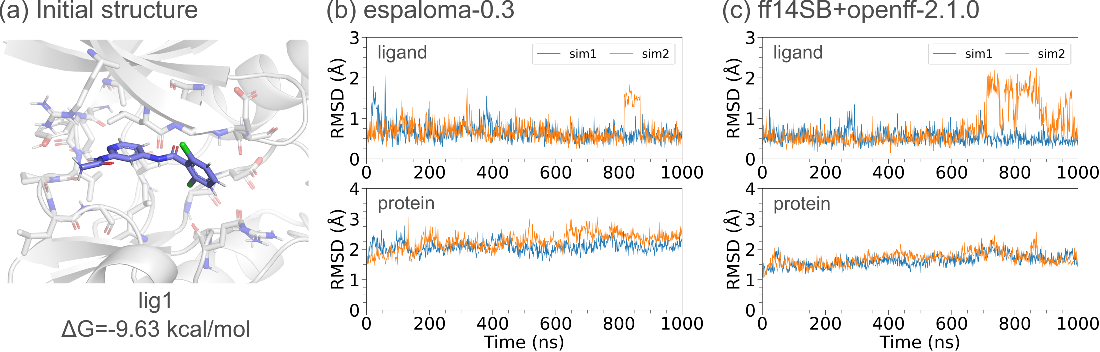}
    \caption{\label{fig:tyk2_vanilla}
    \textbf{espaloma-0.3 is robust and capable of stable long-time MD simulation for the Tyk2 protein-ligand complex system.}
    Multiple one microsecond of MD simulations were conducted on the Tyk2 protein-ligand complex system to explore the stability of \texttt{espaloma-0.3}. (a) We show the initial structure of Tyk2 complexed with ligand \#1. Two protein-ligand complex MD simulations were performed using (b) \texttt{espaloma-0.3} to self-consistently parametrize both the protein and ligand, and (c) \texttt{openff-2.1.0} and Amber ff14SB to parametrize the ligand and protein, respectively.
    The root-mean square deviation (RMSD) profile of the heavy ligand atoms and protein C${\alpha}$ atoms are reported over the one microsecond MD simulation. The trajectories were aligned with respect to the binding pocket residues (within 4 ${\text{Å}}$ from the initial ligand pose) before computing the ligand RMSD. Similarly, the protein C${\alpha}$ atoms excluding the first and last 5 residues, were used to align the protein trajectories before RMSD calculation, with the first and last 5 residues excluded from RMSD computation.
    }
\end{figure}

\end{document}